\documentclass[superscriptaddress,twocolumn,aps,pra,preprintnumbers,notitlepage,showkeys,showpacs]{revtex4-1}
\usepackage{amsmath,amssymb}
\usepackage{lipsum} 
\usepackage[latin9]{inputenc}
\usepackage{graphicx}
\usepackage{dcolumn}
\usepackage{bbold}
\usepackage{bm}
\usepackage[usenames]{xcolor}
\usepackage[colorlinks,bookmarks=false,citecolor=blue,linkcolor=blue,urlcolor=blue,hyperfootnotes=true]{hyperref}
\usepackage{accents}

\begin{document}

\title{DC resistivity near a nematic quantum critical point: Effects of weak disorder and acoustic phonons}

\author{Lucas E. Vieira}
\affiliation{Instituto de Física, Universidade Federal de Goiás, 74.001-970, Goiânia-GO, Brazil}
\author{Vanuildo S. de Carvalho}
\affiliation{Instituto de Física Gleb Wataghin, Unicamp, 13083-859, Campinas-SP, Brazil}
\author{Hermann Freire}
\affiliation{Instituto de Física, Universidade Federal de Goiás, 74.001-970, Goiânia-GO, Brazil}

\date{\today}
\begin{abstract}
We calculate the resistivity associated with an Ising-nematic quantum critical point in the presence of disorder and acoustic phonons in the lattice model. To perform this analysis, we use the memory-matrix transport theory, which has a crucial advantage compared to other methods of not relying on the existence of well-defined quasiparticles in the low-energy effective theory. As a result, we obtain that by including an inevitable interaction between the nematic fluctuations and the elastic degrees of freedom of the lattice (parametrized by the nemato-elastic coupling $\kappa_{\text{latt}}$), the resistivity $\rho(T)$ of the system as a function of temperature obeys a universal scaling form described by $\rho(T)\sim T\ln (1/T)$ at high temperatures, reminiscent of the paradigmatic strange metal regime observed in many strongly correlated compounds. For a window of temperatures comparable with $\kappa^{3/2}_{\text{latt}}\varepsilon_F$ (where $\varepsilon_F$ is the Fermi energy of the microscopic model), the system displays another regime in which the resistivity is consistent with a description in terms of $\rho(T)\sim T^{\alpha}$, where the effective exponent roughly satisfies the inequality $1\lesssim\alpha\lesssim 2$. However, in the low-temperature limit (\textit{i.e.}, $T\ll\kappa^{3/2}_{\text{latt}}\varepsilon_F$), the properties of the quantum critical state change in an important way depending on the types of disorder present in the system: It can either recover a Fermi-liquid-like regime described by $\rho(T)\sim T^2$ or it could exhibit yet another non-Fermi liquid regime characterized by the scaling form $\rho(T)-\rho_0\sim T^2\ln T$ (implying in the latter case that the system would display a Kondo-like upturn in the resistivity). From a broader perspective, our results emphasize the key role played by both phonon and disorder effects in the scenario of nematic quantum criticality and might be fundamental for addressing recent transport experiments in some iron-based superconductors.
\end{abstract}


\maketitle

\section{Introduction} 

There has been an accumulation of evidence in recent years of the likely manifestation of quantum criticality in many important systems such as the high-$T_c$ cuprates \cite{Hinkov-S(2008),Taillefer-N(2010),Ando-PRL(2002),Fujita-S(2014),Matsuda-NP(2017),Kaminski-N(2002),Bourges-CRP(2016)}, the heavy-fermion compounds \cite{Thompson-S(2008)}, and the iron-based superconductors \cite{Kivelson-PRB(2008),Davis-S(2010),Fisher-S(2012),Hussey-N(2019),Coldea-ARCMP(2018),Bohmer-JPCM(2018)} (to name only a few systems). One prominent candidate theory for the emergence of this phenomenon in some of the aforementioned materials is related to the onset of Ising-nematic quantum criticality \cite{Fradkin-ARCMP(2010),Fernandes-NP(2014),Keimer-N(2015)}. For tetragonal environments, this order refers to the spontaneous breaking of a point-group discrete rotation symmetry of the lattice from $C_4$ down to $C_2$, while preserving translation symmetry (i.e., it is an order at momentum $\mathbf{q}=0$). In metallic systems, it can be achieved, e.g., via a Pomeranchuk instability \cite{Pomeranchuk-JETP(1958)} of Landau Fermi-liquid theory in the corresponding charge channel, but it can also appear quite generally in either frustrated or doped incommensurate antiferromagnets once the antiferromagnetism has been destroyed by quantum fluctuations \cite{Read-PRL(1991),*Sachdev-IJMPB(1991)}. 

The original approach to quantum critical phenomena is due to Hertz \cite{Hertz-PRB(1976)} (and later extended by Millis \cite{Millis-PRB(1993)}): It relies on the premise that it is possible to integrate out the fermions to derive an effective action with Landau-damping for the (bosonic) order-parameter field. However, this conventional Hertz-Millis strategy has been challenged due to the fact that it does not treat all excitations of the theory at a given energy scale on equal footing. Despite this cautionary remark, it was thought for some time that such an effective model could be solved exactly \cite{Altshuler-PRB(1994)} at low temperatures in the large-$N_f$ limit, with $N_f$ being the number of fermionic species. Unfortunately, this turned out not to be the case due to strong quantum fluctuation effects that emerge at low energies. Indeed, the problem was declared open again after the work of Lee \cite{Lee-PRB(2009)} on a closely related critical Fermi surface state with the $N_f$ fermions coupled to an emergent U(1) gauge field.

\begin{figure*}[t]
\centering \includegraphics[width=0.44\linewidth]{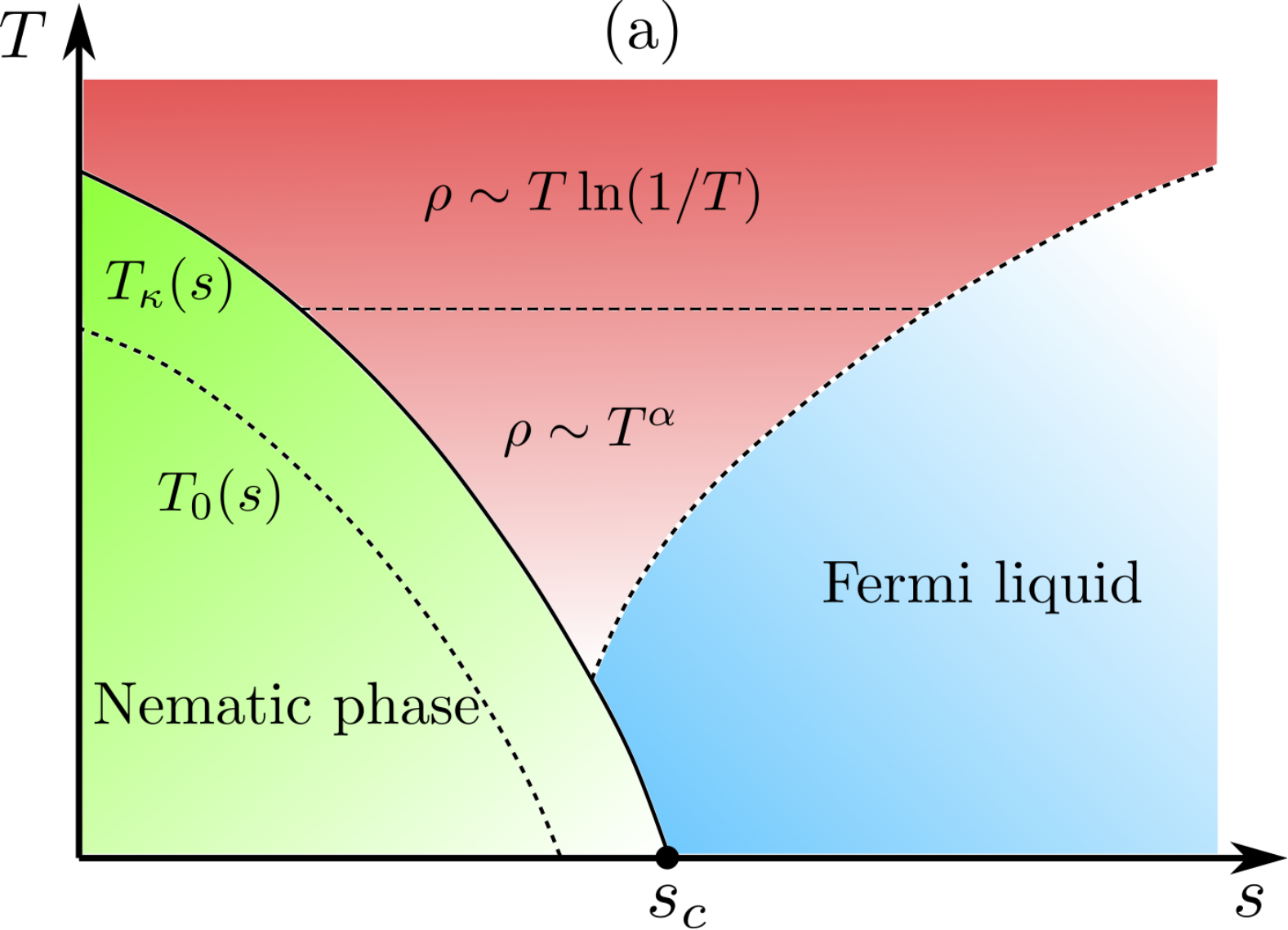}\hfill{}\includegraphics[width=0.44\linewidth]{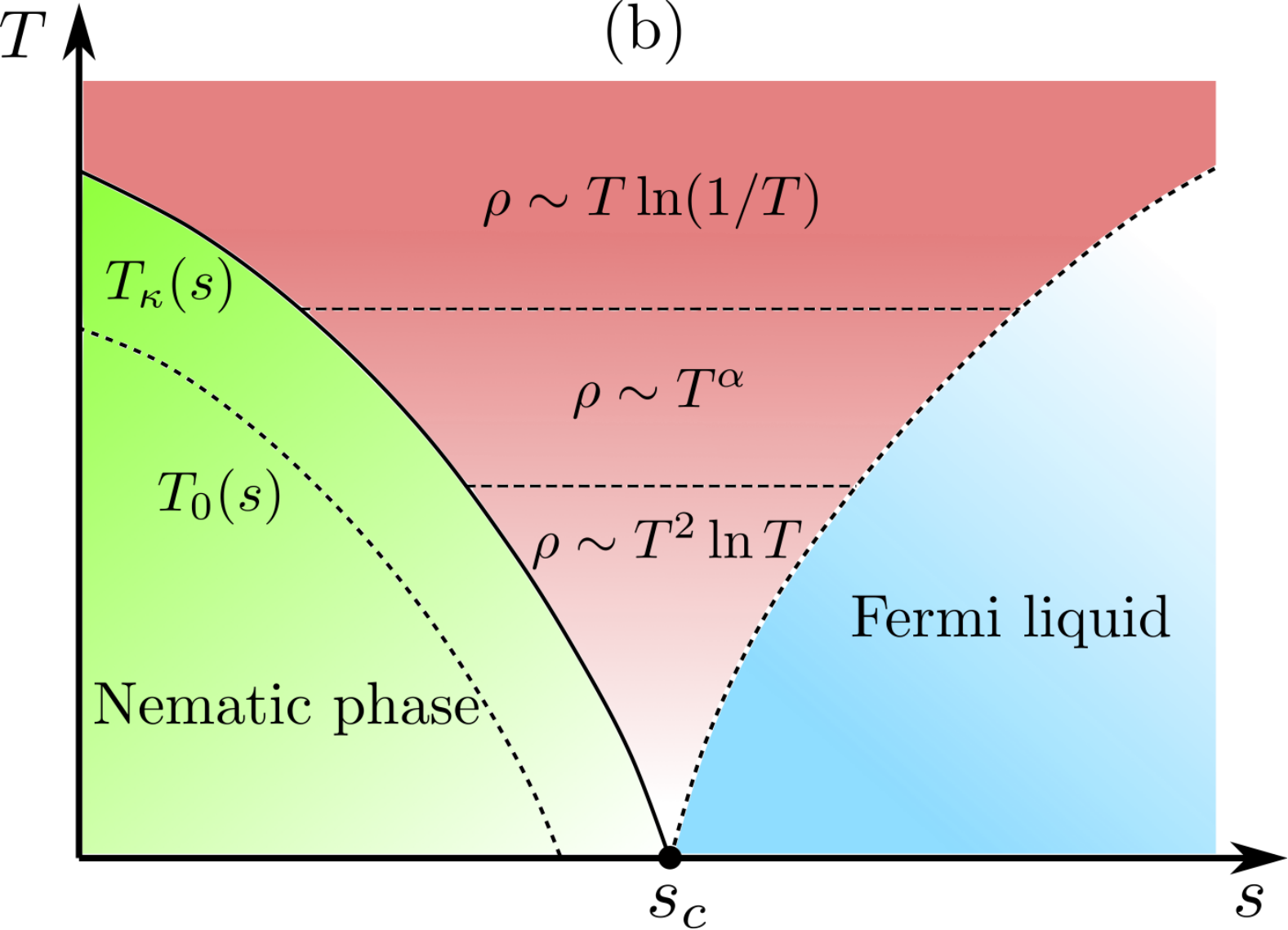}
\caption{Schematic phase diagrams of temperature $T$ versus the tuning parameter $s$ obtained in the present work: (a) represents the case in which only the random potential $V_0$ coupling is included in the theory, and (b) stands for the case in which both random potential $V_0$ and the random field $h_0$ are included. In both phase diagrams, the resistivity exhibits a $T\ln(1/T)$ dependence at high temperatures. For a finite range of intermediate temperatures, the resistivity is consistent with a description in terms of $\rho\sim T^{\alpha}$, with the effective exponent $\alpha$ roughly satisfying the inequality $1\lesssim\alpha\lesssim 2$ in both scenarios. At the lowest temperatures, the system evolves into either a Fermi-liquid-like regime for diagram (a) or to yet another non-Fermi liquid regime characterized by $\rho-\rho_0\sim T^2\ln T$ in diagram (b). The transition temperature to the nematic phase $T_\kappa(s)$ scales as $T_\kappa(s)\sim |\Delta s| \ln(1/|\Delta s|)$ for a finite nemato-elastic coupling $\kappa_\text{latt}$ (where $\Delta s\equiv s-s_c$), also for both scenarios. The $T_0(s)$ refers to the nematic transition temperature for $\kappa_\text{latt}=0$. We also note that the phase diagram in Fig. \ref{Fig_Phase_Diagram}(a) agrees well with the phase diagram first obtained in Ref. \cite{Paul-PRL(2017)} by an analysis of the thermodynamic properties of the model. }\label{Fig_Phase_Diagram}
\end{figure*}

Following those important developments, a novel renormalization group (RG) approach up to three-loop order \cite{Metlitski-PRBa(2010)} was put forward by Metlistki and Sachdev to address this difficult but general problem of Ising-nematic quantum criticality arising in two-dimensional metals. Their central conclusion was that although a systematic $1/N_f$ approximation fails close to the quantum critical point (QCP) -- and, for this reason, there is still no fully-controlled RG program for this theory -- they found interesting renormalizations of several parameters and a clear non-Fermi liquid regime, which emerges inside the corresponding quantum critical region. Other analytical works subsequently confirmed this general picture for the nematic quantum critical scenario \cite{Senthil-PRB(2010),Kopietz-PRB(2012),Metzner-PRB(2015)}. On the other hand, from a numerical perspective, there have been also some recent advances in simulations of this problem using determinantal quantum Monte-Carlo (QMC) methods, where the authors in Refs. \cite{Berg-S(2012),Lederer-PNAS(2017)} confirmed some results obtained previously by the aforementioned analytical methods. Moreover, those numerical works have provided further support to the existence of a non-Fermi liquid phase inside the quantum critical regime at intermediate temperatures and the emergence of a high-$T_c$ superconducting dome in the phase diagram as the temperature is lowered, which have some similarities with the experiment situation \cite{Lederer-PNAS(2017)}. Unfortunately, those QMC simulations are restricted to small lattices and not-too-low temperatures, such that many properties specifically related to the QCP of the model are still not easily accessible.

Transport theories of metallic phases possessing strong quantum critical nematic fluctuations are also of great interest nowadays. Indeed, the resistivity is one of the simplest quantities to be measured experimentally in order to characterize a variety of strongly correlated phases. In this context, a widely employed method for computing non-equilibrium properties is the semi-classical Boltzmann-equation approach \cite{Ziman-OUP(1960)}. In the Hertz-Millis scenario, this method considers the scattering of the low-energy fermions off the order-parameter bosons, which are assumed to be both in equilibrium and to provide a bath that effectively acts to degrade the fermionic total momentum. For the Ising-nematic quantum critical model in two dimensions, this method predicts that the resistivity $\rho(T)$ in the corresponding non-Fermi liquid phase should obey a power-law described by $\rho(T)\sim T^{4/3}$, as a result of  considering impurity scattering as the main mechanism for momentum relaxation and by including also multiple bands to avoid special ``one-dimensional'' geometrical cancellation effects \cite{Maslov-PRL(2011),Pal-LJPTS(2012)}. On the other hand, it has been generally acknowledged that a conventional Boltzmann-equation approach could potentially fail for the Ising-nematic quantum critical problem because of two reasons: Firstly, the quasiparticle excitations are not well-defined in the corresponding non-Fermi liquid that emerges in the quantum critical region of the phase diagram. Secondly, since the $1/N_f$ approximation turns out not to be controlled for this problem of nematic quantum criticality at low temperatures \cite{Metlitski-PRBa(2010)}, the notable Prange-Kadanoff reasoning \cite{Prange-PR(1964)} for the validity of the Boltzmann equation may not work for the present case. Therefore, an alternative technique to calculate transport properties could be used in order to avoid those aforementioned issues.

In this work, we will apply one possible alternative transport theory to the Ising-nematic quantum critical region in two-dimensional metals that can avoid the potential difficulties highlighted above associated with this problem. As the source for momentum relaxation in the present theory, we will focus here on disorder effects as the main contribution, following the study by Ref. \cite{Hartnoll-PRB(2014)} also in the context of a nematic quantum critical theory. Then, we will proceed to include in the calculation of the corresponding transport coefficient important new effects associated with phonon scattering via a nemato-elastic coupling \cite{Xu-PRB(2009),Schmalian-PRL(2010),Dagotto-PRL(2013),Schmalian-PRB(2016),Paul-PRL(2017),Paul-PRB(2017),Carvalho-PRB(2019)}, which from an RG point of view turns out to be as relevant as disorder. The technique we will adopt here will be the Mori-Zwanzig memory matrix approach \cite{Forster-HFBSCF(1975)}, which has the advantage of not being based on the existence of well-defined quasiparticle excitations at low energies in the system \cite{Hartnoll-PRB(2013),Berg-PRB(2019)}. It has been employed in recent years by several researchers of the field in many different contexts, such as, e.g., one-dimensional Luttinger liquids \cite{Rosch-PRL(2000)}, two-dimensional quantum critical metals \cite{Hartnoll-PRB(2014),Patel-PRB(2014),Lucas-PRB(2015),Freire-AP(2017),*Freire-EPL(2017),Freire-EPL(2018)}, a critical (spinon) two-dimensional Fermi surface model coupled to an emergent U(1) gauge field \cite{Freire-AP(2014)}, and holographic quantum matter \cite{Zaanen-CUP(2015),Sachdev-MIT(2018)}. Here, we will focus on the calculation of the resistivity of the corresponding electronic phase associated with Ising-nematic quantum critical fluctuations and its interplay with both disorder and acoustic phonons. As a consequence, we will be able to investigate this transport coefficient as a function of the nemato-elastic interaction, which emerges inevitably in any realistic nematic QCP model defined on a lattice (see also Figs. \ref{Fig_Phase_Diagram}(a) and \ref{Fig_Phase_Diagram}(b) for an overview of the two possible scenarios obtained in the present work).

Therefore, this paper is structured as follows: In Sec. \ref{Sec_Model}, we define the Ising-nematic quantum critical model with nemato-elastic coupling that will be our starting point for the analysis of the corresponding transport properties. Then, in Sec. \ref{Sec_MMatrix} we will briefly explain the Mori-Zwanzig memory-matrix technique that will be used throughout this work. The analytical and the numerical results that emerge from the equations obtained within this method will be presented in Secs. \ref{Sec_TMass} and \ref{Sec_NemRes}. Lastly, we will end this paper with a summary concerning the present investigation.

\section{Model}\label{Sec_Model} 

We start from a standard model that consists of a two-dimensional tetragonal electronic system described by $N_f$ species of fermions coupled to bosonic scalar fields, which represent the Ising-nematic correlations. Those nematic fluctuations are in turn also coupled to the elastic degrees of the freedom of the underlying lattice. The corresponding Hamiltonian in $\mathbf{k}$-space is therefore given by $H=H_{\text{el-nem}}+H_{\text{nem-latt}}$. The first part of the Hamiltonian ($H_{\text{el-nem}}$) that describes the coupling between the electrons and the nematic fluctuations is given by
\begin{align}\label{H1}
&H_{\text{el-nem}} =\sum_{\mathbf{k},\sigma}\xi_{\mathbf{k}} \psi^{\dagger}_{\mathbf{k}\sigma}\psi_{\mathbf{k}\sigma}+\frac{N_f}{2}\sum_{\mathbf{k}} s\, \phi_{\mathbf{k}}^2+\frac{1}{2}\sum_{\mathbf{k}} \pi_{\mathbf{k}}^2\nonumber\\
&+\frac{g_\text{nem}}{\sqrt{\nu_0}}\sum_{\mathbf{k},\mathbf{q},\sigma}V_{\mathbf{k,q}}\psi^{\dagger}_{\mathbf{k}+\mathbf{q}/2,\sigma}\psi_{\mathbf{k}+\mathbf{q}/2,\sigma}\phi_{\mathbf{q}},
\end{align}

\noindent where $\psi^{\dagger}_{\mathbf{k},\sigma}$ and $\psi_{\mathbf{k},\sigma}$, respectively, creates and annihilates electrons with momentum $\mathbf{k}$ and spin projection $\sigma\in\{\uparrow,\downarrow\}$, $\xi_{\mathbf{k}}=\varepsilon_{\mathbf{k}}-\mu$ is the energy dispersion relative to the chemical potential of the system, $s$ is the ``distance'' of the model to the nematic QCP, $\pi_{\mathbf{k}}$ is the momentum operator conjugated to $\phi_{\mathbf{k}}$, $g_{\text{nem}}$ is the nematic interaction that couples the order-parameter field to the electrons, and $V_{\mathbf{k,q}}=2[\cos(k_x)+\cos(q_x)-\cos(k_y)-\cos(q_y)]$ is the corresponding $d$-wave form factor. We introduced the density of states $\nu_0$ in the definition of the nematic interaction only for convenience later on in our calculations. Due to this redefinition, it is the ratio $(g_\text{nem}/\sqrt{\nu_0})$ that has units of energy. The above critical model describes the effects of an electronic structural quantum phase transition from tetragonal ($C_4$) to orthorhombic ($C_2$) symmetry in the system as a function of the control parameter described by $s$.

As for the second part of the Hamiltonian ($H_{\text{nem-latt}}$), which refers to the important coupling between the nematic quantum fluctuations and the elastic degrees of freedom (acoustic phonons) of the underlying two-dimensional lattice, we write
\begin{equation}\label{H2}
H_{\text{nem-latt}} =\frac{1}{2}\sum_{\mathbf{k}\neq 0}\textit{\textbf{u}}^{\dagger}_{\mathbf{k}}\cdot\mathcal{M}_{\mathbf{k}}\cdot\textit{\textbf{u}}_{\mathbf{k}}+ig_{\text{latt}}\sum_{\mathbf{k}\neq 0} ({\textbf{a}}_{\mathbf{k}}\cdot\textit{\textbf{u}}_{\mathbf{k}})\phi_{\mathbf{-k}},
\end{equation}

\noindent where $\textit{\textbf{u}}_{\mathbf{k}}$ is the Fourier transform of the displacement vector, ${\textbf{a}}_{\mathbf{k}}=(k_x,-k_y,0)$ is a two-dimensional vector, $g_{\text{latt}}$ is the nemato-elastic coupling, and $\mathcal{M}_{\mathbf{k}}$ is the so-called dynamical matrix \cite{Paul-PRL(2017),Carvalho-PRB(2019),Schutt-PRB(2018)}. If we constrain the displacement vector $\textit{\textbf{u}}_{\mathbf{k}}$ to lie effectively in two spatial dimensions, the dynamical matrix $\mathcal{M}_{\mathbf{k}}$ can be simplified to
\begin{equation}
\mathcal{M}_{\mathbf{k}}=
  \left( {\begin{array}{cc}
   C_{11} k_x^2+C_{66}k_y^2 & (C_{12}+C_{66})k_x k_y\\
   (C_{12}+C_{66})k_x k_y & C_{66} k_x^2+C_{11}k_y^2 \\
  \end{array} } \right),
\end{equation}
where the constants $C_{ij}$ refer to the elastic constants in the Voigt notation (see, e.g., Ref. \cite{Landau-PP(1970)}) for a system possessing initially tetragonal symmetry \cite{Cowley-PRB(1976)}.

Regarding the role of the nemato-elastic coupling in the present theory, we will follow a treatment similar to the one explained in Refs. \cite{Paul-PRL(2017),Carvalho-PRB(2019),Schutt-PRB(2018)}. We first project the displacement vector onto the basis of the polarization vectors of the phonons and diagonalize the dynamical matrix according to $\mathcal{M}_{\mathbf{k}}\cdot\mathbf{\hat{e}_{\mu}}(\mathbf{k})=\varrho\omega_{\mu}^2\mathbf{\hat{e}_{\mu}}(\mathbf{k})$, where $\varrho$ is the mass density of the ions of the underlying lattice. Then, by integrating out the phonon degrees of freedom, the nematic degrees of freedom become described by the following renormalized propagator $D^{-1}(\mathbf{k},i\Omega_n)=D^{-1}_0(\mathbf{k},i\Omega_n)-\Pi_{\text{latt}}(\mathbf{k},i\Omega_n)$, where
\begin{equation}
\Pi_{\text{latt}}(\mathbf{k},i\Omega_n)=g_{\text{latt}}^2\sum_{\mu}\frac{|{\textbf{a}}_{\mathbf{k}}\cdot \mathbf{\hat{e}_{\mu}}(\mathbf{k})|^2}{\omega_{\mu}^2(\mathbf{k})+\Omega_n^2}.
\end{equation}

\begin{figure}[t]
\includegraphics[width=2.9in]{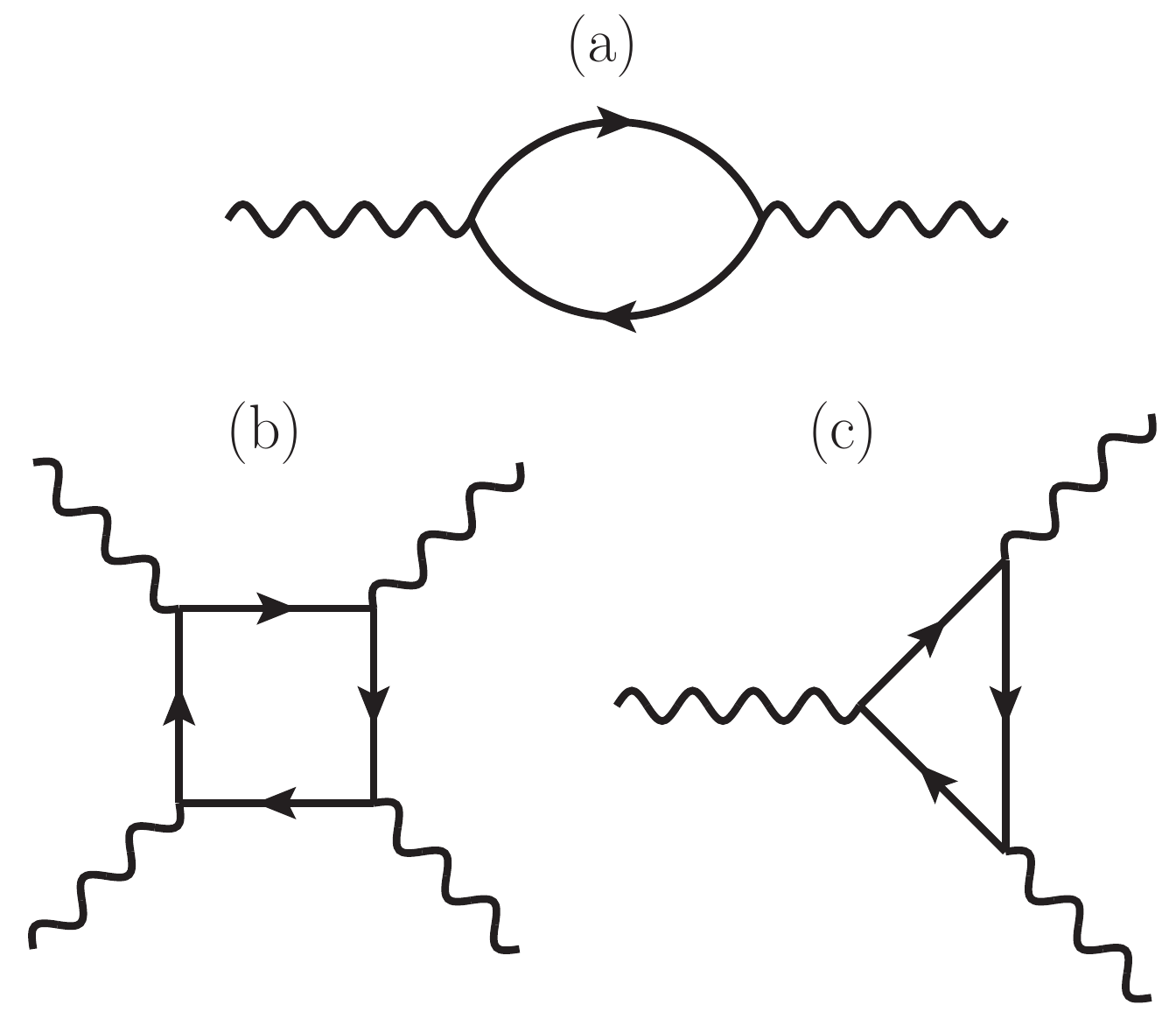}
\caption{Feynman diagrams for (a) the polarizability $\Pi_0$, (b) the bosonic vertex correction $\Gamma_4$, and (c) the three-legged bosonic vertex $\Gamma_3$, respectively. The solid lines stand for the fermionic propagators and
the wavy lines refer to external legs associated with bosonic fields.}\label{Fig_Gamma_Bubbles}
\end{figure}

\noindent By defining the angle $\theta=\tan^{-1}(k_y/k_x)$, we obtain that the eigenvalues of $\mathcal{M}_{\mathbf{k}}$ can be written as $\omega_{\pm}(\mathbf{k})=v_{\pm}(\theta)|\mathbf{k}|$, with the sound velocities given by
\begin{equation}
v_{\pm}(\theta)=\frac{1}{\sqrt{2\varrho}}\sqrt{\gamma_1\pm\sqrt{\frac{\gamma_2^2+\gamma_3^2}{2}+\frac{\gamma_2^2-\gamma_3^2}{2}\cos(4\theta)}},
\end{equation}
where the $\gamma$'s are given in terms of the elastic constants by $\gamma_1=C_{11}+C_{66}$, $\gamma_2=C_{11}-C_{66}$, and $\gamma_3=C_{12}+C_{66}$.

From the above expressions, we can now calculate the ``distance'' to the nematic quantum critical point $\nu^{-1}_0s=D^{-1}_{\text{nem}}(0,0)$ that straightforwardly evaluates to
\begin{equation}
s(\theta)=s_0- \frac{2g_{\text{latt}}^2\nu_0}{\gamma_1^2-F^2(\theta)}\big[\gamma_1+\gamma_3-(\gamma_2+\gamma_3)\cos^2(2\theta)\big],
\end{equation}
where we have defined the function $F(\theta)=\sqrt{\frac{1}{2}(\gamma_2^2+\gamma_3^2)+\frac{1}{2}(\gamma_2^2-\gamma_3^2)\cos(4\theta)}$. By considering the experimental relevant case in which $(C_{11}-C_{12})/2=C_{66}$ that is valid only for the tetragonal phase (we point out that this condition will be always met in our work, because we will be always in the symmetric phase of the present model), this implies that $\gamma_2=\gamma_3$. Therefore, the ``distance'' $s$ to the nematic quantum critical point becomes finally given by
\begin{equation}\label{s_theta}
s(\theta)=\Delta s+\lambda_{\text{latt}}\cos^2(2\theta),
\end{equation}
where $\Delta s\equiv s-s_{c}$, with $s_c=\frac{g^2_\text{latt}\nu_0}{C_{66}}$ and $\lambda_{\text{latt}}=\frac{s_c}{2}\left(1+\frac{C_{12}}{C_{11}}\right)$. At the nematic QCP, the tuning parameter of course vanishes, i.e., $\Delta s=0$. As a consequence, we can see that not only the coupling of the nematic fluctuations produces a shift in the distance to the corresponding QCP, but this shift clearly becomes direction-selective as a function of $\theta$. In other words, there are special directions, in which $\theta_n=(2n+1)\pi/4$ (for integer $n$) that clearly generate no shift in $s(\theta)$ (we also note that these points correspond to the so-called cold spots in the model).

\begin{figure}[t]
\includegraphics[width=3.2in]{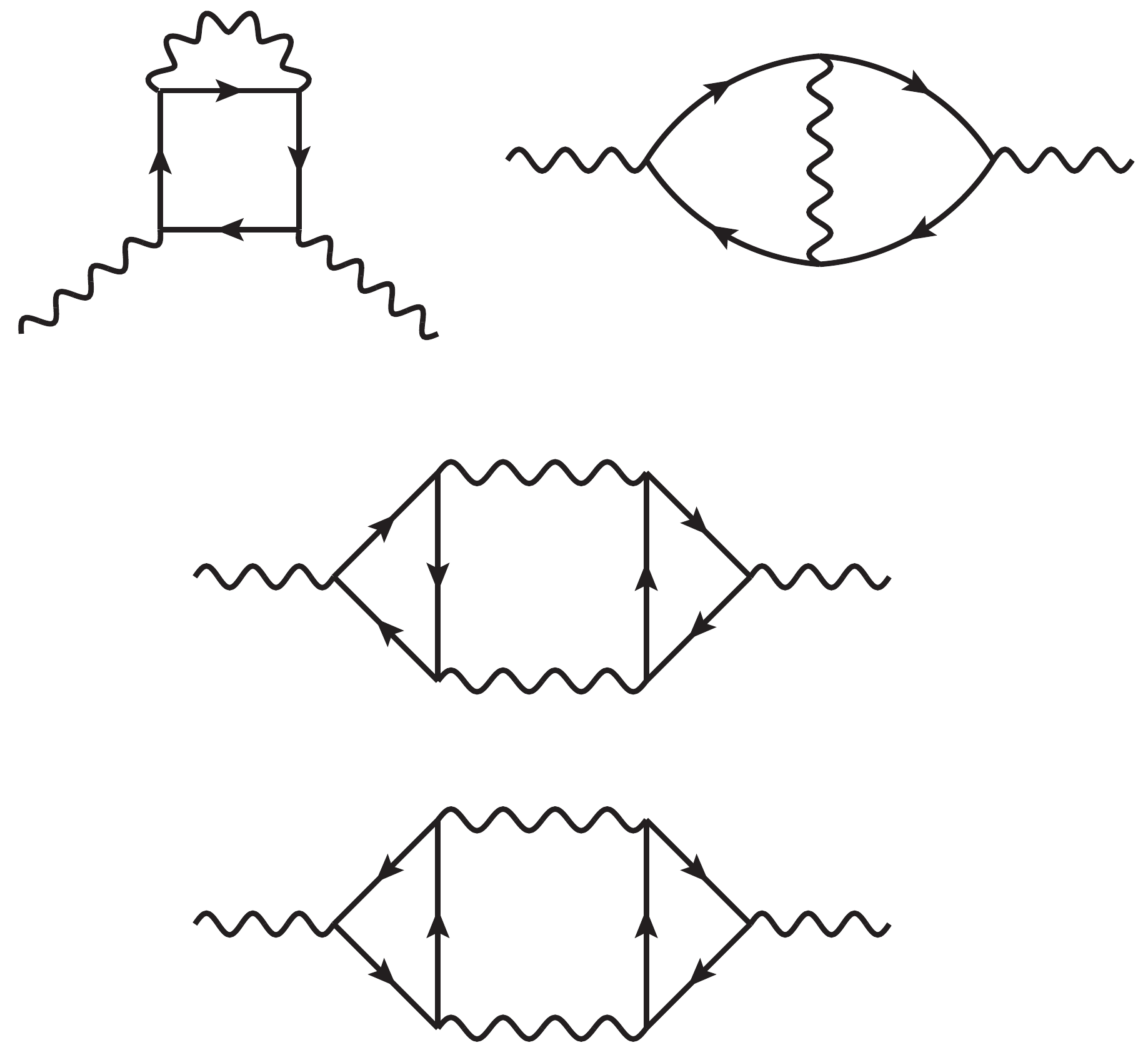}
\caption{The Feynman diagrams for the bosonic self-energy at order $1/N_f$.}\label{Fig_Bos_Self_Energy}
\end{figure}

We are now ready to define the propagator for the bosonic nematic fluctuations as a result of the nemato-elastic interaction described above. This will be one of the central quantities to be utilized in the present work. By assuming that the underlying Fermi surface is of circular shape and focussing on scattering processes that take place on the the vicinity of the Fermi surface, we obtain for $|\Omega_n|\ll v_F k$ that the ``bare'' bosonic propagator is
\begin{widetext}
\begin{equation}\label{Eq_Prop_Zero}
D^{-1}_0(\mathbf{k},i\Omega_n)=\nu^{-1}_0\bigg[\Delta s+k^2+\lambda_\text{latt}\cos^2(2\theta)+\mathcal{B}\cos^2(2\theta)\frac{|\Omega_n|}{v_Fk}+\mathcal{C}\sin^2(2\theta)\left(\frac{\Omega_n}{v_Fk}\right)^2\bigg],
\end{equation}
\end{widetext}
\noindent where $\mathcal{B}$ and $\mathcal{C}$ are usually temperature-independent (non-universal) constants that for the present microscopic model are given by $\mathcal{B}=\nu_0g^2_\text{nem}$ and $\mathcal{C}=4\nu_0g^2_\text{nem}$. Note that we also included the irrelevant term in the bosonic propagator that is multiplied by the prefactor $\mathcal{C}$, which will be important for some calculations that will appear in this work (see the Appendices \ref{Nem_Mass_Equation}--\ref{Chi_JP}).

To proceed further, we will then follow an analysis similar to that in Ref. \cite{Hartnoll-PRB(2014)} in order to calculate the renormalized nematic mass $m^2(T)$ as a function of temperature at the present QCP. In this way, the structure of the corresponding renormalized bosonic propagator along with the temperature-dependent mass-term (e.g., for the case in which $\Delta s=0$) becomes 
\begin{widetext}
\begin{equation}\label{Eq_Prop_Temp}
D^{-1}(\mathbf{k},i\Omega_n)=\nu^{-1}_0\bigg[k^2+\lambda_\text{latt}\cos^2(2\theta)+\mathcal{B}\cos^2(2\theta)\frac{|\Omega_n|}{v_Fk}+\mathcal{C}\sin^2(2\theta)\left(\frac{\Omega_n}{v_Fk}\right)^2+m^2(T)\bigg].
\end{equation}
\end{widetext}
At this point, we may now calculate the renormalized nematic mass $m^2(T)$. In order to do that, we will use a standard $1/N_f$ expansion. According to recent theoretical developments \cite{Xu-PRB(2009),Paul-PRL(2017)}, this is expected to be a valid approximation at low temperatures for the present quantum critical model with the nemato-elastic coupling included. This latter property is related to the fact that the quantum critical theory defined by Eqs. \eqref{H1} and \eqref{H2} describing the Ising-nematic fluctuations interacting with acoustic phonons is governed by a Gaussian fixed point in the clean limit. Therefore, according to the Hertz-Millis strategy, we may for this case integrate out the fermions to derive an effective theory for the nematic order-parameter bosons. As a consequence, the effective action describing the nematic fluctuations truncated at fourth-order in an expansion in powers of the order-parameter field finally becomes
\begin{align}
&\frac{S_{\phi}}{N_f}=\frac{1}{2}\int_{K}D_0^{-1}(K)|\phi(K)|^2+\frac{1}{3}\prod_{i=1}^{3}\int_{K_i}\delta(\sum_{i}K_i)\nonumber\\
&\times \Gamma_ 3(K_1,K_2,K_3)\,\phi(K_1)\phi(K_2)\phi(K_3)+\frac{1}{4}\prod_{i=1}^{4}\int_{K_i}\delta(\sum_{i}K_i)\nonumber\\
&\times \Gamma_ 4(K_1,K_2,K_3,K_4)\,\phi(K_1)\phi(K_2)\phi(K_3)\phi(K_4),\nonumber\\
\end{align}
where $K_i=(\omega_i,\mathbf{k}_i)$ and $\int_{K}(\cdots)=T\sum_{\omega_n}\int\frac{d^2 \mathbf{k}}{(2\pi)^2}(\cdots)$. 

By calculating the corrections at the order $1/N_f$ in the model (see the corresponding Feynman diagrams in Figs. \ref{Fig_Gamma_Bubbles} and \ref{Fig_Bos_Self_Energy}), it can be shown that these contributions are given by
\begin{align}\label{mass}
&m^2(T)=D_0^{-1}(0)+\frac{U}{N_f}\int_{K}D(K)\nonumber\\
&+\frac{2}{N_f}\int_{K}[\Gamma_3(K,-K,0)D(K)]^2,
\end{align}
where the coupling $U$ was defined as $U=2\Gamma_4(K,-K,0,0)+\Gamma_4(K,0,-K,0)$, and $\Gamma_4(K,-K,0,0)$ and $\Gamma_4(K,0,-K,0)$ are given, respectively, by
\begin{align}
&\Gamma_4(K,-K,0,0)=\frac{g_{\text{nem}}^4}{\nu^2_0}T\sum_{\nu_n}\int_\mathbf{q} V^2_{\mathbf{q}+\mathbf{k}/2,\mathbf{q}-\mathbf{k}/2} V^2_{\mathbf{q}-\mathbf{k}/2,\mathbf{q}-\mathbf{k}/2} \nonumber\\
&\times G_0^3(\mathbf{q}+\mathbf{k}/2,i\nu_n+i\omega_n)G_0(\mathbf{q}-\mathbf{k}/2,i\nu_n),\label{Gamma4a}\\
&\Gamma_4(K,0,-K,0)=\frac{g_{\text{nem}}^4}{\nu^2_0}T\sum_{\nu_n}\int_\mathbf{q} V^2_{\mathbf{q}+\mathbf{k}/2,\mathbf{q}-\mathbf{k}/2} V_{\mathbf{q}-\mathbf{k}/2,\mathbf{q}-\mathbf{k}/2} \nonumber\\
&\times V_{\mathbf{q}+\mathbf{k}/2,\mathbf{q}+\mathbf{k}/2} \,G_0^2(\mathbf{q}+\mathbf{k}/2,i\nu_n+i\omega_n)G_0^2(\mathbf{q}-\mathbf{k}/2,i\nu_n).\label{Gamma4b}
\end{align}
As for $\Gamma_3(K,-K,0)$, it evaluates to
\begin{align}
&\Gamma_3(K,-K,0)=\frac{g_{\text{nem}}^4}{\nu^2_0}T\sum_{\nu_n}\int_\mathbf{q} V^2_{\mathbf{q}+\mathbf{k}/2,\mathbf{q}-\mathbf{k}/2} V_{\mathbf{q}-\mathbf{k}/2,\mathbf{q}-\mathbf{k}/2} \nonumber\\
&\times G_0^2(\mathbf{q}+\mathbf{k}/2,i\nu_n+i\omega_n)G_0(\mathbf{q}-\mathbf{k}/2,i\nu_n).\label{Gamma3}
\end{align}
In all the above equations, the non-interacting fermionic Green's function is given by $G_0(\mathbf{k},i\omega_n)=1/(i\omega_n-\xi_{\mathbf{k}})$ for the energy dispersion $\xi_{\mathbf{k}}$ and $\int_{\mathbf{q}}(\cdots)=\int\frac{d^2 \mathbf{q}}{(2\pi)^2}(\cdots)$. Moreover, since we are assuming a second-order nematic QCP in the present model, we will only consider here the case in which $U>0$.

At the nematic QCP (i.e., $\Delta s=0$), we must have that $m^2(T=0)=0$. This is related to fact that the theory should be manifestly gauge-invariant at the corresponding nematic QCP. Therefore, we should rewrite the Eq. (\ref{mass}) in a more convenient way
\begin{align}\label{mass2}
m^2(T)=&\Delta s+\frac{U}{N_f}\int_{\mathbf{k}}\left[T\sum_{\omega_n}D(\mathbf{k},i\omega_n)\hspace{-0.05cm}
-\hspace{-0.15cm}\int_{\omega}D_0(\mathbf{k},\omega)\bigg|_{\Delta s=0}\right]\nonumber\\
&+\frac{2}{N_f}\int_{K}[\Gamma_3(K,-K,0)D(K)]^2,
\end{align}
\noindent where  $\int_{\omega}(\cdots)=\int^{\infty}_{-\infty}\frac{d \omega}{2\pi}(\cdots)$. The above equation turns out to be a self-consistent equation for $m^2(T)$. We will then proceed to solve it after the coming section, which in turn contains a brief explanation of the methodology that we will employ in the present work.

\section{Memory-Matrix Method}\label{Sec_MMatrix} 

The method that we will adopt in this work to calculate transport properties of the nematic quantum critical model will be the Mori-Zwanzig memory-matrix formalism \cite{Forster-HFBSCF(1975),Zaanen-CUP(2015),Sachdev-MIT(2018)}. As advertised previously, this technique has the advantage of not relying on the existence of well-defined quasiparticle excitations. According to this formalism, the matrix of ``generalized'' conductivities can be written as follows
\begin{equation}\label{sigma}
{\sigma}(\omega,T)=\frac{{\chi}^R(T)}{({M}-i\omega{{\chi}^R(T)})[{{\chi}^R}(T)]^{-1}},
\end{equation}
where $\chi_{R}(T)$ stands for the static retarded susceptibility matrices representing the overlap of the currents of interest with the so-called nearly-conserved operators in the system. These susceptibility matrices can in turn be expressed as 
\begin{align}\label{susc}
\chi_{{A}{B}}(i\omega,T)=\int_{0}^{1/T}d\tau e^{i\omega\tau}\langle T_{\tau}{A}^{\dagger}(\tau)B(0)\rangle,
\end{align}
where the corresponding retarded susceptibility is obtained by performing the analytical continuation $i\omega\rightarrow\omega +i0^{+}$. In addition, $\langle(\cdots)\rangle$ represents the grand-canonical average, $T_{\tau}$ is the imaginary-time ordering operator and the volume has been set to unity, for simplicity. The memory matrix denoted by ${M}$ can then be obtained from the formally exact expression 
\begin{equation}\label{memory}
{M}_{\mathcal{O}_j \mathcal{O}_\ell}(T)=\int_{0}^{1/T}d\tau\left\langle \dot{\mathcal{O}}^{\dagger}_j(0)Q\frac{i}{\omega-Q{L}Q}Q\dot{\mathcal{O}}_\ell(i\tau)\right\rangle,
\end{equation}

\noindent where the super-operator ${L}$ is the Liouville operator, which is conventionally defined as ${L}\,\mathcal{O}_j=[H,\mathcal{O}_j]=-i\dot{\mathcal{O}}_j$, $H$ is the Hamiltonian of the model, and $\mathcal{O}_j$ refers in principle to an arbitrary operator. For practical reasons though, this method becomes much more useful if the operators $\mathcal{O}_j$ turn out to be either conserved or nearly conserved (i.e., slowly varying) in the transport theory. Consequently, the memory matrix will necessarily become a small quantity, which will allow its calculation via perturbative means. The $Q$ denotes another super-operator that acts on the operators and projects out of a formal space spanned by all the conserved (or nearly conserved) operators in the system. 

At this point, it is also useful to define the Lagrangian density of the system corresponding to Eqs. \eqref{H1} and \eqref{H2} with the acoustic phonons being already integrated out in order to apply the Mori-Zwanzig memory-matrix formalism. In position space denoted by $\mathbf{r}$ and imaginary time $\tau$, it can be written as
\begin{align}\label{Lagrangian}
\mathcal{L}&=\sum_{\sigma}{\psi}^{\dagger}_{\sigma}\left(\partial_{\tau}-\frac{\nabla_{\mathbf{r}}^2}{2m_f}-\mu\right)\psi_{\sigma}+\frac{s}{2}N_f{\phi}^2+\frac{\epsilon}{2}(\partial_{\tau}{\phi})^2\nonumber\\
&-\frac{g_\text{nem}}{\sqrt{\nu_0}}\phi\sum_{\sigma}\left ({\psi}^{\dagger}_{\sigma}[(\partial_x^2-\partial_y^2)\psi_{\sigma}]+[(\partial_x^2-\partial_y^2){\psi}^{\dagger}_{\sigma}]\psi_{\sigma}\right).
\end{align}

\noindent where $m_f$ is the mass of the fermions. In the above equation, we have also included the bosonic kinetic term (multiplied by the parameter $\epsilon$) in order to obtain the total momentum of the model. Since this particular bosonic contribution turns out to be irrelevant at low energies due to Landau damping, we could set $\epsilon\rightarrow 0$ in the end of the present transport coefficient computation, but actually the final result turns out to be independent of $\epsilon$ \cite{Hartnoll-PRB(2014)}. 

Within the framework of Noether's theorem, the fact that the Lagrangian of the system is invariant under continuous space translations and global U(1) symmetry implies that the total momentum $\mathbf{P}$ and the electric current $\mathbf{J}$ are conserved at the classical level. These quantities are, respectively, given by
\begin{align}\label{momentum_current}
\mathbf{P}&=\frac{i}{2}\sum_{\sigma} \left[({\nabla{\psi}}^{\dagger}_{\sigma}\,\psi_{\sigma}-{\psi}^{\dagger}_{\sigma}{\nabla\psi}_{\sigma})+\epsilon\nabla{\phi}\,\partial_{\tau}{\phi}\right],\\
{J_x}&=i\left(\frac{1}{2m_f}+2\frac{g_\text{nem}}{\sqrt{\nu_0}}\phi\right)\sum_{\sigma} (\partial_x\psi^{\dagger}_{\sigma}\,\psi_{\sigma}-\psi^{\dagger}_{\sigma}\partial_x\psi_{\sigma}),
\\
{J_y}&=i\left(\frac{1}{2m_f}-2\frac{g_\text{nem}}{\sqrt{\nu_0}}\phi\right)\sum_{\sigma} (\partial_y\psi^{\dagger}_{\sigma}\,\psi_{\sigma}-\psi^{\dagger}_{\sigma}\partial_y\psi_{\sigma}),
\end{align}

\noindent where the total momentum $\mathbf{P}$ has both the contribution from the electrons and also the bosonic (drag) term (the latter corresponding to the Ising-nematic fluctuations). As for the electric current $\mathbf{J}$, we clearly obtain two distinct contributions: one corresponding to the non-interacting part and another that includes the interaction $g_{\text{nem}}$ between the electron and the nematic fluctuation. This shows that such an interaction affects the transport of charge in the system as well.

\begin{figure}[t]
\includegraphics[width=1.0\linewidth]{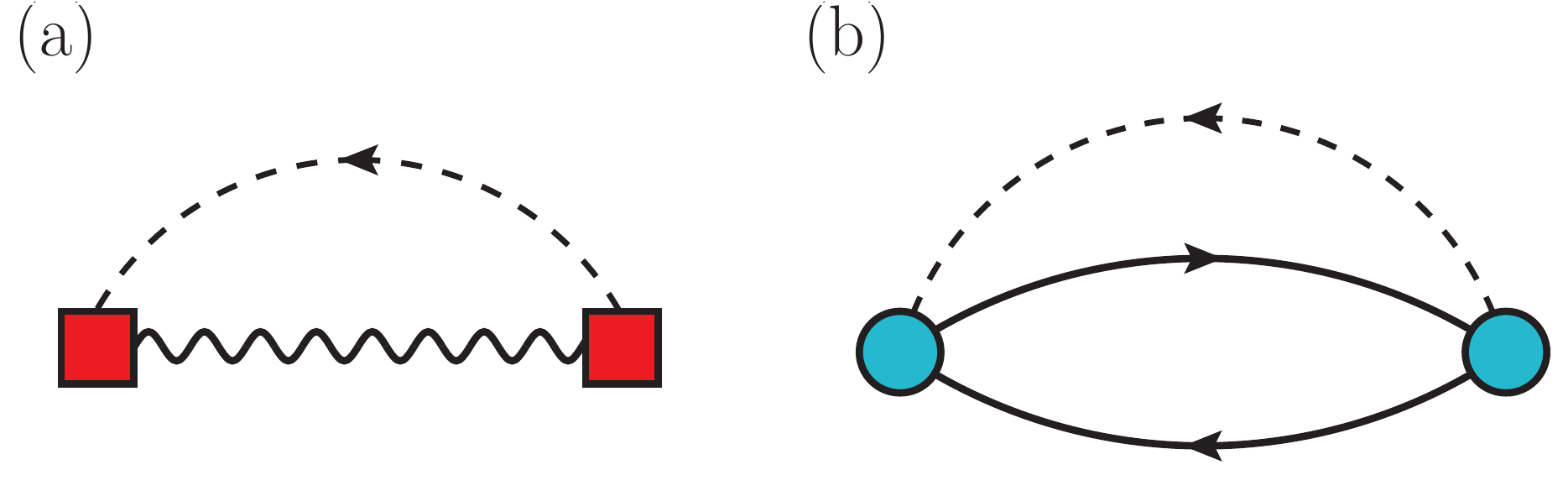}
\caption{Feynman diagrams for the calculation of the leading contributions to the memory matrix in the present theory: (a) the random-field-disorder contribution where the red squares represent the $h_0$ coupling, and (b) the random potential contribution coupled to the fermions where the blue circles stand for the $V_0$ coupling. The
dashed lines in both diagrams refer to the impurity lines and carry only internal momentum and external energy $\omega$.}\label{MM_Diagram}
\end{figure}

In the present context of a nemato-elastic quantum critical model, our starting point will be the following: Since the electrons, the acoustic phonons and the nematic fluctuations are strongly interacting with one another at low temperatures, they are naturally able to exchange momentum very rapidly in the model. Therefore, we will adopt here the point of view (also put forward in many other works in the literature \cite{Hartnoll-PRB(2013),Hartnoll-PRB(2014),Patel-PRB(2014),Lucas-PRB(2015),Freire-AP(2017),*Freire-EPL(2017),Freire-EPL(2018)}) that for the computation of the resistivity in the strange metal regime, the momentum-conserving interactions -- that are naturally dominant in the model and result in the corresponding non-Fermi-liquid physics -- are also the fastest processes leading to local equilibrium. In other words, the total momentum $\mathbf{P}$ turns out to be the only slowly-varying operator in the present system, and the corresponding resistivity is then controlled by scattering processes that do not conserve the total momentum. Two possibilities arise in this case: impurity scattering and/or Umklapp. On general grounds, Umklapp processes depend on subtle issues such as the shape of the Fermi surface of the model and its distance to the Brillouin zone boundary, the number of electrons (or holes) in the conduction band, etc. At very low temperatures, these latter processes are expected to become, in some physical situations, less efficient in relaxing the total momenta of the system. For this reason, we will choose to focus here (only for the sake of simplicity) on impurity scattering as the mechanism for momentum relaxation in the present problem and leave the interesting analysis of Umklapp scattering for a future investigation. The present transport property calculation is thus based on the so-called Peierls scenario \cite{Peierls(1991)} for momentum relaxation and should be contrasted with the other well-known Bloch scenario, in which the drag effect of the total momentum does not play an important role. 

As a result, we will include here two types of impurity terms in the present model that couple, respectively, to the electrons and the nematic fluctuations in the Lagrangian density of Eq. \eqref{Lagrangian}, i.e.
\begin{eqnarray}\label{Eq_Imp_Lag}
\mathcal{L}_\text{imp}&=&\sum_{\sigma}V(\mathbf{r})\psi^{\dagger}_{\sigma}(\mathbf{r})\psi_{\sigma}(\mathbf{r})+h(\mathbf{r}){\phi}(\mathbf{r}),
\end{eqnarray}
which should satisfy Gaussian disorder averages: $\langle\langle V(\mathbf{r}) \rangle\rangle=\langle\langle h(\mathbf{r}) \rangle\rangle=0$,
$\langle\langle V(\mathbf{r})V(\mathbf{r'}) \rangle\rangle=V_0^2\delta^2(\mathbf{r}-\mathbf{r'})$, and $\langle\langle h(\mathbf{r})h(\mathbf{r'}) \rangle\rangle=h_0^2\delta^2(\mathbf{r}-\mathbf{r'})$, where $V_0$ is a random potential for the electron field and the parameter $h_0$ is the so-called random field disorder that couples to the nematic fluctuation $\phi$. Both contributions were considered in an earlier work \cite{Hartnoll-PRB(2014)} for the case of $\lambda_{\text{latt}}=0$. 

Assuming that the couplings to the random potential $V_0$ and to the random field disorder $h_0$ are both small, we now perform a perturbative calculation of the memory matrix in the present model. Since it can be shown that the equation for $\dot{\mathbf{P}}=i[H,\mathbf{P}]$ is of order
linear in both $V_0$ and $h_0$, the leading contribution to the memory matrix will depend quadratically on those couplings. Therefore, the leading contribution to the Liouville operator defined after Eq. \eqref{memory} will be given by its non-interacting value (i.e., ${L}\approx {L}_0$) and, for the same reason, the grand-canonical averages should also be calculated in terms of the non-interacting Hamiltonian of the present system, i.e. $\langle(\cdots)\rangle_0$. As a result, the projector $Q$ in Eq. \eqref{memory} will not contribute within this approximation. Thus, to leading order, the conductivity in Eq. \eqref{sigma} reduces to (for transport along the $x$-direction):
$\sigma_{dc}=\lim_{\omega\rightarrow 0} \chi_{JP}^2\langle\dot{{P_x}}\frac{i}{\omega-{L}_0}\dot{{P_x}}\rangle_0^{-1}$. We then obtain \cite{Forster-HFBSCF(1975),Patel-PRB(2014),Hartnoll-PRB(2014)}
\begin{align}\label{Eq_Conductivity_Form}
{\sigma_{dc}}&=\lim_{\omega\rightarrow 0}\frac{\chi_{JP}^{2}}{\frac{1}{\omega}\int_{0}^{\infty}dt e^{i\omega t}\langle[\dot{{P_x}}(t),\dot{{P_x}}(0)]\rangle_0},\\
\rho(T)&=\frac{1}{\sigma_{dc}}=\frac{1}{\chi_{JP}^{2}}\lim_{\omega\rightarrow 0}\frac{\text{Im}\,G^R_{\dot{{P_x}}\dot{{P_x}}}(\omega,T)}{\omega},
\end{align}
where $G^R_{\dot{{P_x}} \dot{{P_x}}}(\omega,T)=\langle \dot{{P_x}}(\omega) \dot{{P_x}}(-\omega)\rangle_0$ is the corresponding retarded Green's function in the Matsubara formalism. Consequently, the leading contributions to the resistivity of the present model (see Fig. \ref{MM_Diagram}) are given by
\begin{align}\label{Eq_Res_Form}
\rho(T)&=\frac{1}{\chi_{JP}^2} \lim_{\omega\rightarrow 0} \int\frac{d^2 \mathbf{k}}{(2\pi)^2}k^2 \cos^2(\theta_{\mathbf{k}}-\varphi) \bigg[h_0^2\frac{\operatorname{Im}D_R(\mathbf{k},\omega)}{\omega}\nonumber\\
&+V_0^2 \frac{\operatorname{Im}\Pi_R(\mathbf{k},\omega)}{\omega}\bigg],
\end{align}
where $\theta_{\mathbf{k}}$ is the polar angle associated with the wavevector $\mathbf{k}=k(\cos\theta_{\mathbf{k}},\sin\theta_{\mathbf{k}})$ and we are considering, for the sake of generality, that the transport of electric current is along an arbitrary angle $\varphi$. Besides, $D_R(\mathbf{k},\omega)=D(\mathbf{k},i\omega\rightarrow\omega+i0^{+})$ is the retarded nematic propagator, $\Pi_R(\mathbf{k},\omega)=\Pi(\mathbf{k},i\omega\rightarrow\omega+i0^{+})$ is the retarded free-fermion polarizability with $\Pi(\mathbf{k},i\omega_n)=-T\sum_{\nu_n}\int\frac{d^2 \mathbf{q}}{(2\pi)^2} G_{0}(\mathbf{k+q},i\omega_n+i\nu_n)G_{0}(\mathbf{q},i\nu_n)$, and $\chi_{JP}$ is the susceptibility given by Eq. \eqref{susc}. Finally, we point out that from Eq. \eqref{sigma} we can still write $\sigma_{dc}=\Gamma^{-1}\chi_{JP}$, where $\Gamma$ is the ``matrix of relaxation rates'', which generalizes the concept of transport scattering rate $(1/\tau_{\text{trans}})$ in the present theory.

To keep our discussion simple at this point, we will first compute here the effect of the random-field disorder contribution to the resistivity and postpone the analysis of the $V_0$ contribution to the resistivity to a later section in this work. In this way, by calculating analytically only the $h^2_0$ term in Eq. \eqref{Eq_Res_Form}, we obtain that it becomes
\begin{align}\label{Eq_Exac_Res}
\rho_{h_0}(T)&=\frac{\mathcal{B}\nu_0h_0^2}{8\pi v_F\chi_{JP}^2\lambda_{\text{latt}}\sqrt{\lambda_{\text{latt}}+m^2(T)}}\bigg\{[\lambda_{\text{latt}}+m^2(T)] \nonumber\\
&\times E\left[\frac{\lambda_{\text{latt}}}{\lambda_{\text{latt}}+m^2(T)}\right]-m^2(T) K\left[\frac{\lambda_{\text{latt}}}{\lambda_{\text{latt}}+m^2(T)}\right]\bigg\},
\end{align}

\noindent where $E(x)$ and $K(x)$ are elliptic functions defined, respectively, according to $E(x)\equiv\int_{0}^{\pi/2}[1-x\sin^2(\vartheta)]^{1/2}d\vartheta$ and $K(x)\equiv\int_{0}^{\pi/2}[1-x\sin^2(\vartheta)]^{-1/2}d\vartheta$. To further substantiate our result, we provide in Appendix \ref{Fermionic_Greens_function} an alternative (but approximate) derivation of the above expression for the resistivity using a traditional fermionic Green's function method. We also point out that if we set $\lambda_{\text{latt}}=0$ (i.e., with no phonons included), we naturally recover the result obtained earlier in Ref. \cite{Hartnoll-PRB(2014)}.

\begin{figure*}[t]
\centering \includegraphics[width=0.47\linewidth]{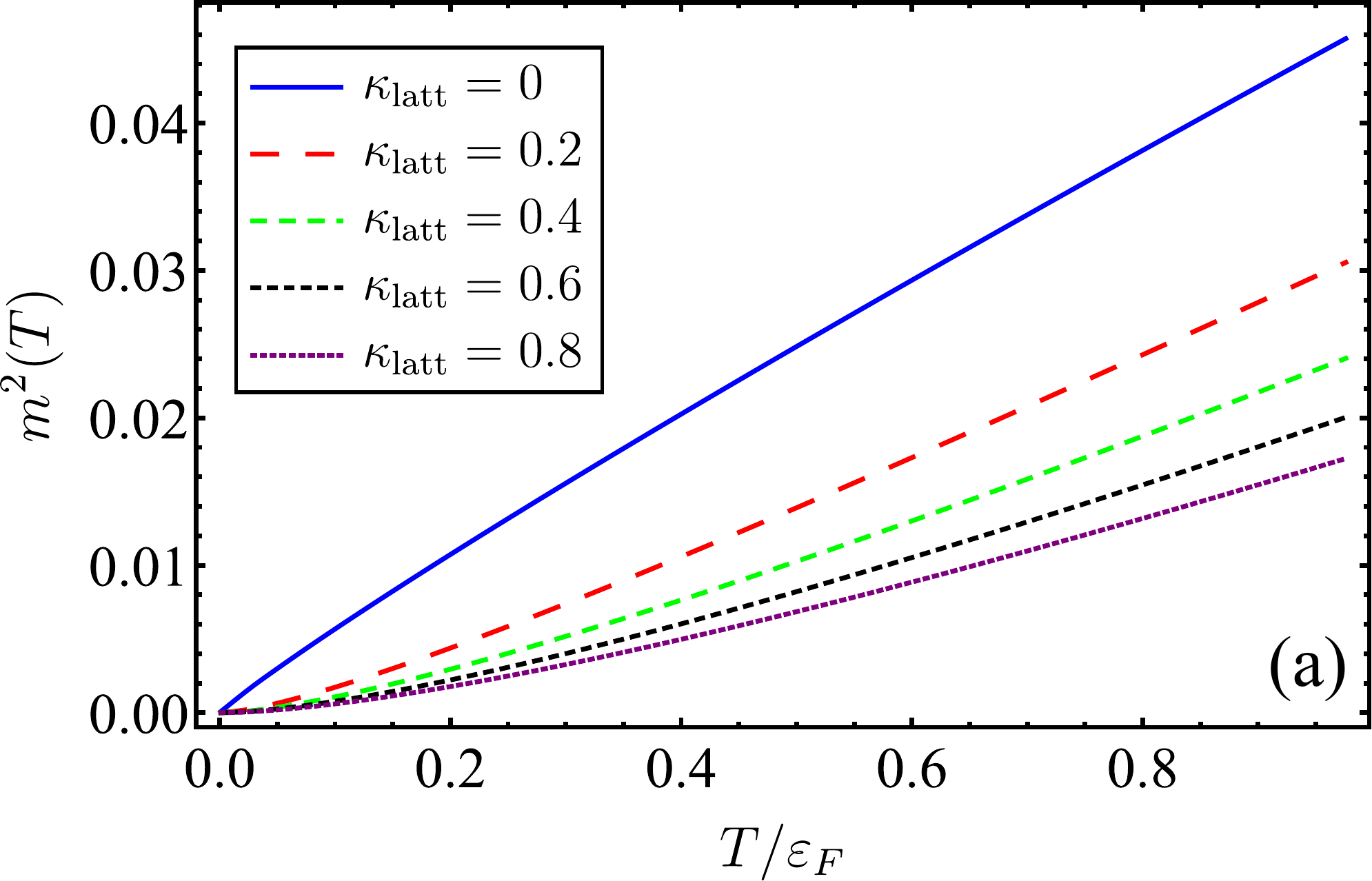}\hfill{}\includegraphics[width=0.47\linewidth]{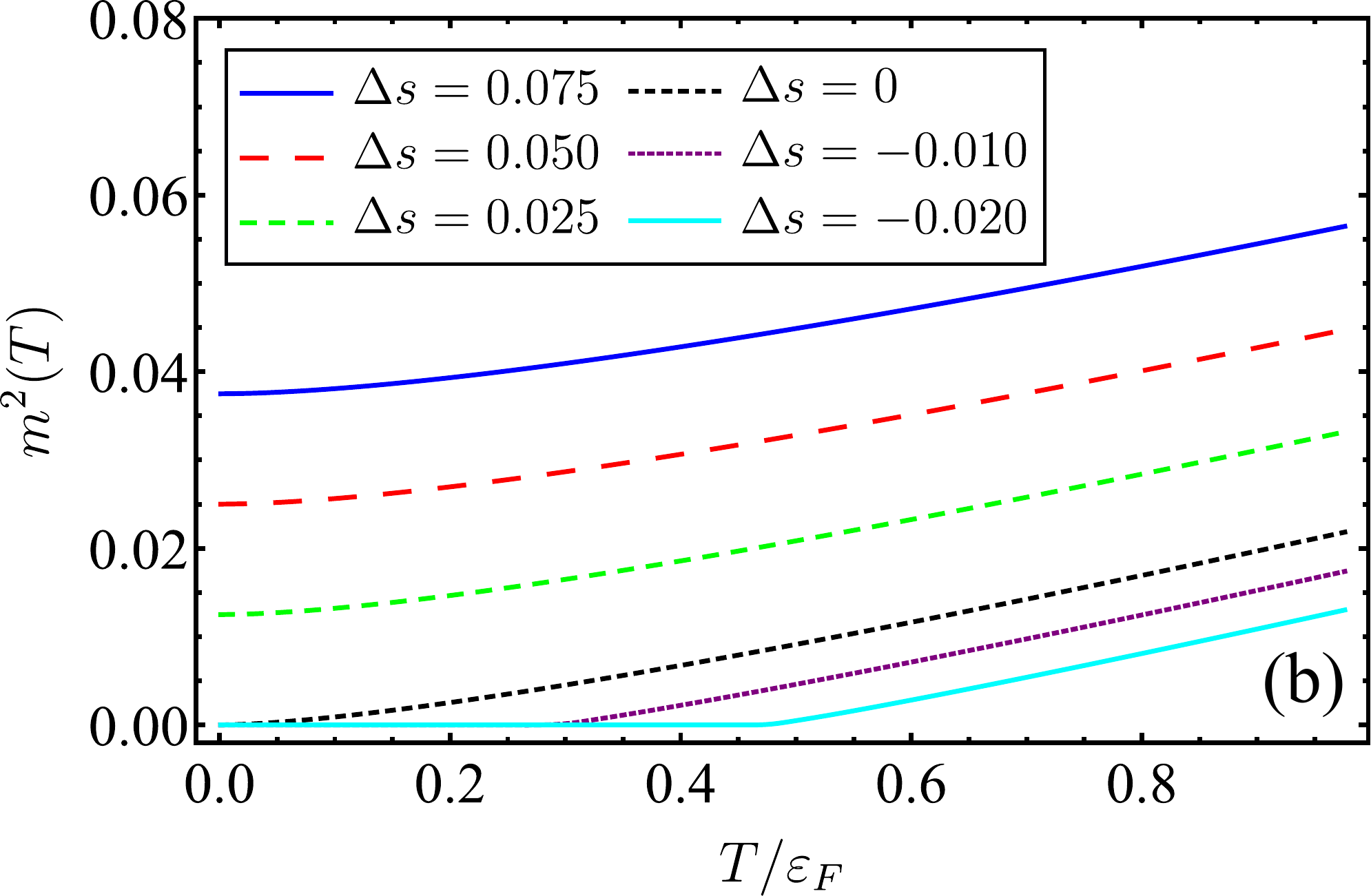}
\caption{(a) Nematic effective mass (in units of $U$) as a function of temperature $T$ for several choices of the nemato-elastic coupling $\kappa_\text{latt}$ at $\Delta{s}=0$. (b) Same quantity (in units of $U$) as a function of $T$ for several choices of the tuning parameter $\Delta{s}$ (for this latter plot, we set the nemato-elastic coupling equal to $\kappa_\text{latt}=0.5$ and, for $\Delta s <0$, we stop our numerical calculation at $T= T_\kappa(s)$, which marks the onset of Ising-nematic order).}\label{Nematc_Mass_Sol}
\end{figure*}

In order to have an idea of the low-temperature dependence of the resistivity, we can expand it to first order in the ratio $m^2(T)/\lambda_\text{latt}$. This yields
\begin{align}\label{Eq_Approx_Res}
\rho_{h_0}(T)&\approx \frac{\mathcal{B}\nu_0h_0^2}{8\pi v_F\lambda^{1/2}_\text{latt}\chi_{JP}^2(T)}\bigg\{1+\frac{1}{4}\bigg[1-4\ln(2)\nonumber\\
&\left.+\ln\left(\frac{m^2(T)}{\lambda_\text{latt}}\right)\right]\frac{m^2(T)}{\lambda_\text{latt}}\bigg\}. 
\end{align}

\noindent In other words, as the temperature approaches the absolute zero, the resistivity $\rho_{h_0}(T)$ turns out to be logarithmically dependent on $m^2(T)/\lambda_\text{latt}$. However, it will eventually saturate at a finite value $\rho^{(0)}_{h_0}=\frac{\mathcal{B}\nu_0h_0^2}{8\pi v_F\lambda^{1/2}_\text{latt}\chi^{(0)2}_{JP}}$, where $\chi^{(0)}_{JP}=\chi_{JP}(T\rightarrow 0)$. This behavior makes also evident that, at least at the nematic QCP, $\rho_{h_0}(T)$ will approach such a finite value from below as the temperature is lowered, since in this case $m^2(T)$ goes to zero. However, to further exploit the dependence of $\rho(T)$ on the temperature and the nemato-elastic coupling, we first have to determine $m^2(T)$. This is the subject of the next section and it will be one of the central results of the present work.

\section{Temperature dependence of the nematic mass}\label{Sec_TMass}

To begin with, if we consider the coupling of the nematic order parameter with the acoustic phonons, it can be shown that the integral that contains $\Gamma_3(K,-K,0)$ in Eq. \eqref{mass2} is naturally free of infrared divergences (as expected, since we will always be in the symmetric phase of the model). However, the same integral also contributes with a non-universal (i.e., ultraviolet-dependent) term to the present critical theory, which can be neglected since we are interested only in the low-energy limit of the model. Therefore, the Eq. \eqref{mass2} for the nematic effective mass in the large-$N_f$ limit becomes
\begin{widetext}
\begin{align}\label{Eq_Nem_Mass_Def}
m^2(T)=&\;\Delta s+\frac{U}{N_f}\int\frac{d^2\mathbf{k}}{(2\pi)^2}\bigg[T\sum_{\omega_n}\frac{\nu_0}{k^2+\lambda_\text{latt}\cos^2(2\theta)+\mathcal{B}\cos^2(2\theta)|\omega_n|/(v_Fk)+\mathcal{C}\sin^2(2\theta)\omega^2_n/(v_Fk)^2+m^2(T)}\nonumber\\
&-\int\frac{d\omega}{2\pi}\frac{\nu_0}{k^2+\lambda_\text{latt}\cos^2(2\theta)+\mathcal{B}\cos^2(2\theta)|\omega|/(v_Fk)+\mathcal{C}\sin^2(2\theta)\omega^2/(v_Fk)^2}\bigg],
\end{align}
\end{widetext}

\noindent where we substituted the nematic propagators defined in Eqs. \eqref{Eq_Prop_Zero} and \eqref{Eq_Prop_Temp}. All the details regarding the evaluation of the momentum-frequency integrals and the Matsubara sum on the right-hand side of Eq. \eqref{Eq_Nem_Mass_Def} are described in Appendix \ref{Nem_Mass_Equation}. By following our analysis developed in that Appendix, we obtain that the equation for $m^2(T)$ evaluates to
\begin{align}\label{Eq_Nem_Mass}
m^2(T)=&\;\frac{\Delta s}{1+\gamma}+\frac{\nu_0UT}{2\pi (1+\gamma)N_f}\nonumber\\
&\times\Xi\bigg[\frac{m(T)}{(2\pi \mathcal{B}T/v_F)^{1/3}},\frac{\lambda^{1/2}_\text{latt}}{(2\pi \mathcal{B}T/v_F)^{1/3}}\bigg],
\end{align}
where the function $\Xi(x,\tau)$ is defined in Eq. \eqref{Eq_Xi} and the parameter $\gamma$ refers to
\begin{align}
\gamma&=\frac{v_F\nu_0U}{4\pi^3\sqrt{\mathcal{C}}N_f}\int^{\Lambda\sqrt{\mathcal{C}}/\mathcal{B}}_0d\rho\int^{2\pi}_0d\theta\int^{\infty}_0dv\nonumber\\
&\times\frac{\rho^5}{[\rho^4+4\kappa_\text{latt}\rho^2\cos^2(2\theta)+\rho v\cos^2(2\theta)+v^2\sin^2(2\theta)]^2}.
\end{align}
Here, $\Lambda$ denotes an ultraviolet momentum cutoff and $\kappa_\text{latt}\equiv\lambda_\text{latt}/(\nu_0g^2_\text{nem})=\frac{1}{2C_{66}}\left(1+\frac{C_{12}}{C_{11}}\right)\frac{g^2_\text{latt}}{g^2_\text{nem}}$ is a dimensionless parameter that measures the strength of the lattice coupling with respect to the nematic interaction \cite{Carvalho-PRB(2019)}. The numerical solution of the Eq. \eqref{Eq_Nem_Mass} is displayed in Fig. \ref{Nematc_Mass_Sol}. In Fig. \ref{Nematc_Mass_Sol}(a), we can see that if the effective nemato-elastic coupling $\kappa_{\text{latt}}$ is equal to zero (i.e., with no phonons included), the effective mass at the nematic QCP obeys the following scaling form $m^2(T)\sim T\ln(1/T)$ (for details, see Appendix \ref{Asymp_Mass_Sol}), which agrees with Ref. \cite{Hartnoll-PRB(2014)}. As the nemato-elastic coupling becomes finite, we obtain for the first time that this effective mass continues to be described by a $T$-linear contribution at high enough temperatures, but its dependence at the nematic QCP evolves into a $T^2$ behavior at temperatures $T\ll\kappa^{3/2}_\text{latt}\varepsilon_F$ (where $\varepsilon_F=v_Fk_F$ is the Fermi energy), which corresponds to the low-temperature limit. In fact, as described in Appendix \ref{Asymp_Mass_Sol}, we are able to obtain at $s=s_c$ the following analytical result 
\begin{equation}\label{Eq_MT_Nem_Mass}
m^2(T)=\frac{16\pi^2k^2_F}{\eta^2\kappa^2_\text{latt}}W^2_0\left(\frac{\eta}{2}\sqrt{\frac{\alpha\Gamma}{1-\beta\Gamma}}\right)\bigg(\frac{T}{\varepsilon_F}\bigg)^2,
\end{equation}

\noindent where $\alpha\approx0.156$, $\beta\approx1.698$, $\eta\approx1.277$, $W_0(z)$ is the first branch of the Lambert function, and $\Gamma$ is a model parameter given in Eq. \eqref{Eq_Gamma_Par}. Notice that this behavior is only possible when the coupling of the electronic system to the lattice degrees of freedom is kept finite. In addition, we show in Fig. \ref{Nematc_Mass_Sol}(b) that the effect of varying the tuning parameter $\Delta s$ for positive values does not change this $T^2$ contribution at low temperatures.

\begin{figure}[t]
\centering \includegraphics[width=1.0\linewidth]{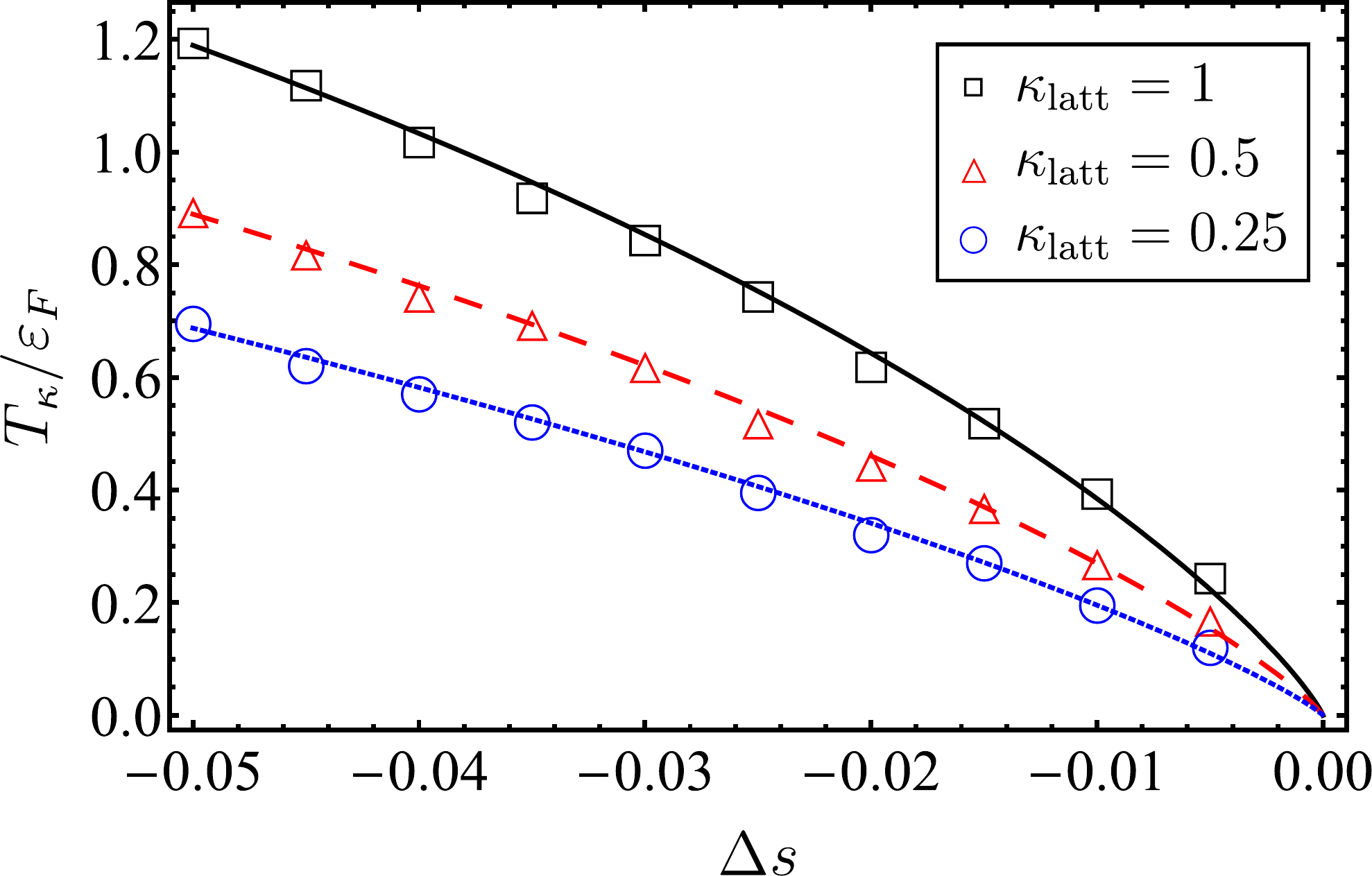}
\caption{The nematic transition temperature $T_\kappa(s)$ versus the tuning parameter $\Delta s<0$ for some choices of $\kappa_\text{latt}$. The lines refer to the best fitting curves described by the family of functions given by $T_\kappa(s)=a_\kappa|\Delta s|\ln(b_\kappa/|\Delta s|$), where $a_\kappa$ and $b_\kappa$ are fitting parameters that depend on the nemato-elastic coupling $\kappa_\text{latt}$.}\label{Critical_Temperature}
\end{figure}

By contrast, in the situation where the control parameter satisfies $\Delta s<0$, the nematic effective mass $m^2(T)$ vanishes for $T= T_\kappa(s)$ (see Fig. \ref{Nematc_Mass_Sol}(b)). In this respect, we would like to emphasize that for $\Delta s <0$ we stop our numerical calculation at $T= T_\kappa(s)$ and never go to lower temperatures, which would imply considering Ising-nematic broken symmetry effects in our present theory. Therefore, our present calculation departing from the tetragonal phase successfully predicts that there is a finite temperature transition, which signals the onset of Ising-nematic order in the model. We have also checked numerically that the dependence of $T_\kappa(s)$ is consistent with the scaling form $T_\kappa(s)\sim |\Delta s|\ln(1/|\Delta s|)$, which is displayed in Fig. \ref{Critical_Temperature}. This scaling form is universal and independent of $\kappa_\text{latt}$ (see also Figs. \ref{Fig_Phase_Diagram}(a) and \ref{Fig_Phase_Diagram}(b) for the qualitative phase diagrams corresponding to the two scenarios obtained in this work). On the other hand, we note that the prefactor associated with the aforementioned functional dependence of  $T_\kappa(s)$ is non-universal and strongly depends on the interaction $\kappa_\text{latt}$. In fact, the higher the coupling $\kappa_\text{latt}$ of the nematic fluctuations to the acoustic phonons the larger the critical temperature associated with the onset of nematic order in the model. This confirms that such an unavoidable coupling to the lattice turns out to be beneficial for the enhancement of nematic fluctuations in the system \cite{Paul-PRL(2017)}.

\section{Contributions to the dc resistivity}\label{Sec_NemRes}

From Eq. \eqref{Eq_Res_Form}, one can see that in order to obtain the full profile of the temperature dependence of $\rho(T)$, one still has to calculate the temperature dependence of $\chi_{JP}$. The details of this calculation are shown in Appendix \ref{Chi_JP}. We will now proceed to discuss the contributions to the resistivity coming from the two types of disorder considered in the present work.

\subsection{Random potential coupled to the fermions}

First, we analyze the random potential contribution (proportional to $V_0^2$ in Eq. \eqref{Eq_Res_Form}) coupled to the fermions. For simplicity, we will consider here only the forward scattering channel, since it is expected to be the most relevant contribution in the present model. Other types of contributions (backscattering, large-angle scattering, Umklapp) will not be considered here, and are left for a future investigation. 

The nematic fluctuations interacting with the fermionic degrees of freedom in a single patch on the Fermi surface will give only subleading corrections to the free fermion polarizability \cite{Metlitski-PRBa(2010),Sur-PRB(2014),Kim-PRB(1994),Stern-PRB(1995)}, whose imaginary part is given by $\operatorname{Im}\Pi_R(\mathbf{k},\omega)=c_b\frac{\omega}{v_Fk}$ with $c_b=m_f/(2\pi)$. As a consequence, by applying that result to our model with a Fermi surface of circular shape, the fermionic contribution to the resistivity yields
\begin{align}\label{Eq_Imp_Fermion}
\rho_{V_0}(T)=c_\Lambda\frac{V_0^2}{\chi^2_{JP}(T)},
\end{align}

\noindent where $c_\Lambda$ is an ultraviolet cutoff-dependent prefactor. This gives a $T$-independent residual resistivity $\rho^{(0)}_{V_0}=c_\Lambda\frac{V^2_0}{\chi^{(0)}_{JP}}$ plus an additional term, from which comes the temperature dependence of $\chi^2_{JP}$. As shown in the Appendix \ref{Chi_JP}, the susceptibility $\chi_{JP}(T)$ in the limit of low temperatures ($T\ll\kappa^{3/2}_\text{latt}\varepsilon_F$) becomes naturally described at $s=s_c$ by
\begin{equation}\label{Eq_LowT_Chi_JP}
\chi_{JP}(T)=\chi^{(0)}_{JP}-\frac{16\pi^2\mathcal{C}_0N_fk^2_F}{\eta^2U\kappa^2_\text{latt}}W^2_0\left(\frac{\eta}{2}\sqrt{\frac{\alpha\Gamma}{1-\beta\Gamma}}\right)\bigg(\frac{T}{\varepsilon_F}\bigg)^2,
\end{equation}
where $\mathcal{C}_0$ is a temperature-independent constant. Consequently, the resistivity in this regime behaves as
\begin{equation}
\frac{\rho_{V_0}(T)-\rho^{(0)}_{V_0}}{\rho^{(0)}_{V_0}}=\frac{32\pi^2\mathcal{C}_0N_fk^2_F}{\eta^2U\chi^{(0)}_{JP}\kappa^2_\text{latt}}W^2_0\left(\frac{\eta}{2}\sqrt{\frac{\alpha\Gamma}{1-\beta\Gamma}}\right)\bigg(\frac{T}{\varepsilon_F}\bigg)^2,
\end{equation}
which suggests that the system exhibits a behavior that resembles a Fermi-liquid regime \footnote{Rigorously speaking, no quenched disorder is necessary in a Fermi liquid to produce a $\rho\sim T^2$ resistivity \cite{Mahan(2000)}. To make the statement about Fermi-liquid behavior in the present model more precise, one needs to investigate also the effect of Umklapp interactions. As mentioned previously, we leave this important analysis for a future work.}. This result is in line with some recent experimental measurements on the compound FeSe$_{0.89}$S$_{0.11}$ \cite{Coldea-NP(2019),Bristow-PRR(2020)}, which show a Fermi-liquid-like dependence of the resistivity as a function of temperature in the vicinity of a putative nematic QCP. In addition, by using the approximation $W_0(x)\approx x$ for the Lambert function, which is valid for small $x$, we find 
\begin{equation}\label{Eq_Asymp_ResV}
\rho_{V_0}(T)-\rho^{(0)}_{V_0}\propto\bigg(\frac{a_1}{\kappa^{3/2}_\text{latt}}+\frac{a_2}{\kappa_\text{latt}}\bigg)\bigg(\frac{T}{\varepsilon_F}\bigg)^2,
\end{equation}
where $a_1$ and $a_2$ are constants defined in terms of parameters of the model. The second term proportional to $\frac{1}{\kappa_\text{latt}}\left(\frac{T}{\varepsilon_F}\right)^2$ in the above expression for $\rho_{V_0}(T)$ was also found in a recent study  \cite{Carvalho-PRB(2019)} of the same quantum critical model by employing the Boltzmann-equation approach to evaluate the resistivity. However, it is very important to point out here that the difference between the result in Ref. \cite{Carvalho-PRB(2019)} and the expression obtained above within the memory matrix approach is the first term on the right-hand side of Eq. \eqref{Eq_Asymp_ResV}, which does not appear in the Boltzmann-equation approach. Since the coupling to the lattice $\kappa_\text{latt}$ is expected to be a small parameter in the model, one may conclude that the first term will dominate over the contribution of the second term. As a result, although the temperature dependence of the resistivity turns out to be the same within both formalisms, the most important contribution to this transport coefficient with respect to  $\kappa_\text{latt}$ is only obtained here within the memory matrix approach.

In the high-temperature limit $\kappa^{3/2}_\text{latt}\varepsilon_F\ll T\ll\varepsilon_F$, the term proportional to the nemato-elastic coupling appearing in Eq. \eqref{Eq_Nem_Mass} becomes irrelevant and the resistivity assumes the form derived in Ref. \cite{Hartnoll-PRB(2014)}, i.e., it becomes nearly $T$-linear with a coefficient proportional to $U$. At intermediate temperatures, it can be verified by the numerical solution of the equation for $m^2(T)$ that there is a window in which the resistivity of the present model is consistent with a description in terms of $\rho(T)\sim T^{\alpha}$, with the effective exponent $\alpha$ satisfying the inequality $1\lesssim\alpha\lesssim 2$. This is due to the fact that $\rho(T)\propto m^2(T)$, when the zero-temperature contribution of $\chi_{JP}(T)$ dominates over $m^2(T)$, as can be seen from the results in Eq. \eqref{Eq_Imp_Fermion} and in Appendix \ref{Chi_JP}. Moreover, this temperature dependence is qualitatively similar to the result obtained in Ref.  \cite{Carvalho-PRB(2019)} by one of us using the Boltzmann-equation approach, although the prefactors multiplying the $T$-dependence turn out to be different in the two formalisms (as we have already explained in the previous paragraph). Such an intermediate regime naturally interpolates between the nearly $T$-linear resistivity that emerges at high temperatures and the $T^2$ behavior which appears at low temperatures (see Fig. \ref{Fig_Phase_Diagram}(a) for the schematic phase diagram obtained within this scenario). We also note that the phase diagram in Fig. \ref{Fig_Phase_Diagram}(a) agrees well with the phase diagram first obtained in Ref. \cite{Paul-PRL(2017)} by an analysis of the thermodynamic properties of the model. Therefore, the results for the resistivity obtained here with the random potential disorder are also in line with recent experimental data reported by Coldea and collaborators \cite{Coldea-NP(2019),Bristow-PRR(2020)}, where it was observed for the FeSe$_{0.89}$S$_{0.11}$ compound under applied external pressure that there is a regime where the corresponding resistivity displays a $T^{\alpha}$-behavior (with a seemingly temperature-dependent $\alpha$ exponent, which also satisfies the inequality $1\lesssim\alpha\lesssim 2$).

On the other hand, in the low-temperature limit, it can be shown via a power-counting scaling analysis that the random potential $V_0$ coupling turns out to be less relevant (from an RG point of view) than the random field disorder $h_0$ coupling. Therefore, at low temperatures, the random-field disorder (if present in the system) will likely dominate the transport properties near the nematic QCP in a model with acoustic phonons (this conclusion also agrees with Ref. \cite{Hartnoll-PRB(2014)} who investigated a nematic quantum critical model, but with no phonons included). For this reason, we now turn our attention to this potentially relevant contribution for the present model in the low-temperature regime.

\subsection{Addition of random-field disorder coupled to the nematic fluctuations}

Here, we analyze (in addition to the contribution of the type of disorder considered in the previous section) the contribution in the present quantum critical model coming from yet another possible type of disorder that some iron-based superconductors could also potentially host in the system -- i.e., the random-field disorder coupled to the nematic order parameter defined in Eq. \eqref{Eq_Imp_Lag}. In the limit of low enough temperatures ($T\ll\kappa^{3/2}_\text{latt}\varepsilon_F$), we obtain from our Eqs. \eqref{Eq_Approx_Res} and \eqref{Eq_LowT_Chi_JP} that the resistivity associated with the nematic quantum critical phase due to the random field disorder in the presence of acoustic phonons is given, to leading order, by
\begin{align}\label{Eq_Res_h0}
\frac{\rho_{h_0}(T)-\rho^{(0)}_{h_0}}{\rho^{(0)}_{h_0}}=\frac{\mathcal{C}_1}{4}\bigg(\frac{T}{\varepsilon_F}\bigg)^2\ln\bigg[\mathcal{C}_1\bigg(\frac{T}{\varepsilon_F}\bigg)^2\bigg]+\mathcal{C}_2\bigg(\frac{T}{\varepsilon_F}\bigg)^2,
\end{align}

\noindent where $\rho^{(0)}_{h_0}$ is, as we have discussed before, a temperature-independent residual resistivity and we have also defined
\begin{align}
\mathcal{C}_1&=\frac{16\pi^2k^2_F}{\eta^2\nu_0g^2_\text{nem}\kappa^3_\text{latt}}W^2_0\left(\frac{\eta}{2}\sqrt{\frac{\alpha\Gamma}{1-\beta\Gamma}}\right),\\
\mathcal{C}_2&=\frac{4\pi^2k^2_F}{\eta^2\kappa^2_\text{latt}}\bigg[\frac{8\mathcal{C}_0N_f}{U\chi^{(0)}_{JP}}-\frac{4\ln(2)-1}{\nu_0g^2_\text{nem}\kappa_\text{latt}}\bigg]W^2_0\left(\frac{\eta}{2}\sqrt{\frac{\alpha\Gamma}{1-\beta\Gamma}}\right).
\end{align}
Consequently, the resistivity in the present situation at low temperatures evolves into an approximate $T^2$-behavior, which one could naively think that a Fermi-liquid-like behavior would also be recovered for this regime as a result of the coupling of the nematic fluctuations to the lattice. However, it is important to point out that, in the low-temperature limit, the logarithmic correction appearing in Eq. \eqref{Eq_Res_h0} will eventually dominate the transport properties of the model for this particular case. This means that in the vicinity of the nematic QCP a non-Fermi-liquid with a Kondo-like upturn is obtained due to the addition of random-field disorder. Despite this, apart from this important difference, other transport properties turn out to be similar to the ones discussed so far: a nearly $T$-linear resistivity at high-temperatures and a regime consistent with $\rho(T)\sim T^{\alpha}$ at intermediate temperatures with the effective exponent $\alpha$ roughly satisfying the inequality $1\lesssim\alpha\lesssim 2$ (see Fig. \ref{Fig_Phase_Diagram}(b) for the schematic phase diagram associated with this second scenario). In this way, our result implies from a broad perspective that the low-temperature behavior associated with Ising-nematic quantum criticality in the presence of acoustic phonons depends in a crucial way on the types of disorder that a given material may host in the system. This could provide a unified explanation of different behaviors in the resistivity observed at low temperatures in many iron-based superconducting compounds available in the literature.

\section{Summary}\label{Sec_Summary}

In this work, we have calculated perturbatively the DC resistivity associated with an Ising-nematic quantum critical point in the presence of both disorder and acoustic phonons in the corresponding lattice model. We have employed the Mori-Zwanzig memory-matrix transport theory, which does not rely on the existence of well-defined quasiparticles in the effective theory. As a result, we have obtained that by including the unavoidable interaction between the nematic fluctuations and the acoustic phonons, those fluctuations clearly become enhanced, and a higher transition critical temperature associated with the onset of nematic order appears. Moreover, the resistivity $\rho(T)$ in the tetragonal phase as a function of temperature becomes described by a universal scaling form given by $\rho(T)\sim T\ln (1/T)$ at high temperatures, reminiscent of the strange metal regime observed in many quantum critical materials. By contrast, for a window of temperatures comparable with $\kappa^{3/2}_{\text{latt}}\varepsilon_F$ (where $\varepsilon_F$ is the Fermi energy of the microscopic model), the system exhibits a regime in which the resistivity is consistent with a description in terms of $\rho(T)\sim T^{\alpha}$, where the effective exponent roughly satisfies the inequality $1\lesssim\alpha\lesssim 2$. Lastly, towards the low-temperature limit, we have shown that the properties of the quantum critical state change in an important way depending on the types of disorder present in the system: It can either recover a Fermi-liquid-like behavior described by $\rho(T)\sim T^2$ for the case where only the random potential $V_0$ coupling is included in the model, or it could display a crossover to yet another non-Fermi liquid phase characterized by the scaling form $\rho(T)-\rho_0\sim T^2\ln T$ due to the effects of the random-field disorder coupling $h_0$ (implying in the latter case that the system would exhibit a Kondo-like upturn in the resistivity). We have argued that our present results could be important for the interpretation, e.g., of recent transport experiments performed in the FeSe$_{0.89}$S$_{0.11}$ compound under applied pressure, and potentially in many other iron-based superconducting compounds as well.

\section*{Acknowledgments}\label{Section_SV}

One of us (H.F.) acknowledges funding from CNPq under Grants No. 405584/2016-4 and No. 310710/2018-9. Partial support from the Fundação de Amparo à Pesquisa do Estado de Goiás (FAPEG) is also greatly appreciated. V.S.deC. acknowledges financial support from FAPESP and Capes under Grants Nos. 2016/05069-7 and 88887.469170/2019-00, respectively.

\appendix

\section{Derivation of the nematic-mass equation}\label{Nem_Mass_Equation}

To evaluate the Matsubara sum and the integrals on the right-hand side of Eq. \eqref{Eq_Nem_Mass_Def}, we follow a similar treatment as devised in Ref. \cite{Hartnoll-PRB(2014)} for a lattice-free model and write Eq. \eqref{Eq_Nem_Mass_Def} as a sum of the two terms
\begin{widetext}
\begin{align}
\mathcal{I}_1=&\int\frac{d^2\mathbf{k}}{(2\pi)^2}\bigg[T\sum_{\omega_n}\frac{\nu_0}{k^2+\lambda_\text{latt}\cos^2(2\theta)+\mathcal{B}\cos^2(2\theta)|\omega_n|/(v_Fk)+\mathcal{C}\sin^2(2\theta)\omega^2_n/(v_Fk)^2+m^2(T)}\nonumber\\
&-\int\frac{d\omega}{2\pi}\frac{\nu_0}{k^2+\lambda_\text{latt}\cos^2(2\theta)+\mathcal{B}\cos^2(2\theta)|\omega|/(v_Fk)+\mathcal{C}\sin^2(2\theta)\omega^2/(v_Fk)^2+m^2(T)}\bigg],\label{Eq_I1}
\end{align}
\begin{align}
\mathcal{I}_2=&\int\frac{d^2\mathbf{k}}{(2\pi)^2}\bigg[\int\frac{d\omega}{2\pi}\frac{\nu_0}{k^2+\lambda_\text{latt}\cos^2(2\theta)+\mathcal{B}\cos^2(2\theta)|\omega|/(v_Fk)+\mathcal{C}\sin^2(2\theta)\omega^2/(v_Fk)^2+m^2(T)}\nonumber\\
&-\int\frac{d\omega}{2\pi}\frac{\nu_0}{k^2+\lambda_\text{latt}\cos^2(2\theta)+\mathcal{B}\cos^2(2\theta)|\omega|/(v_Fk)+\mathcal{C}\sin^2(2\theta)\omega^2/(v_Fk)^2}\bigg],\label{Eq_I2}
\end{align}
\end{widetext}

\noindent where, as explained in the main text, $\lambda_\text{latt}$ is associated with the coupling between the acoustic phonons and the nematic order parameter.

In order to find a convergent expression for $\mathcal{I}_1$, it is not necessary to keep the formally irrelevant $\omega^2$ term in the nematic propagator. Indeed, to prove that, we first recall that the bosonic Matsubara frequency is given by $\omega_n=2\pi nT$ and then define the frequency cutoff $\Omega=2\pi N T$, where $N$ is a very large integer number. Having this in mind, we write $\mathcal{I}_1$ as the limit
\begin{widetext}
\begin{align}
\mathcal{I}_1=&\int\frac{d^2\mathbf{k}}{(2\pi)^2}\lim_{\Omega\rightarrow\infty}\bigg[T\sum^{N}_{n=-N}\frac{\nu_0}{k^2+\lambda_\text{latt}\cos^2(2\theta)+m^2(T)+2\pi\mathcal{B}T\cos^2(2\theta)|n|/(v_Fk)}\nonumber\\
&-\int^{\Omega}_{-\Omega}\frac{d\omega}{2\pi}\frac{\nu_0}{k^2+\lambda_\text{latt}\cos^2(2\theta)+m^2(T)+\mathcal{B}\cos^2(2\theta)|\omega|/(v_Fk)}\bigg].
\end{align}
\end{widetext}

\noindent Next, we can proceed with the evaluation of the Matsubara sum for $N$ finite by utilizing the following mathematical identity
\begin{align}
\sum^N_{n=-N}\frac{1}{a+b|n|}=&\;\frac{2}{b}\bigg[\psi^{(0)}\bigg(\frac{a}{b}+N+1\bigg)-\psi^{(0)}\bigg(\frac{a}{b}+1\bigg)\bigg]\nonumber\\
&+\frac{1}{a}.
\end{align}
Here, the parameters $a$ and $b$ correspond to positive real numbers and $\psi^{(0)}(z)$ denotes the digamma function. Thus, using the set of definitions 
\begin{align}
a&\equiv k^2+\lambda_\text{latt}\cos^2(2\theta)+m^2(T),\\
b&\equiv\frac{2\pi\mathcal{B}T}{v_Fk}\cos^2(2\theta),
\end{align}
one can show that
\begin{align}
\mathcal{I}_1&=\nu_0\int\frac{d^2\mathbf{k}}{(2\pi)^2}\bigg\lbrace \frac{T}{a}+\frac{2T}{b}\lim_{\Omega\rightarrow\infty}\bigg[\psi^{(0)}\bigg(\frac{a}{b}+\frac{\Omega}{2\pi T}+1\bigg)\nonumber\\
&-\ln\bigg(1+\frac{b\Omega}{2\pi aT}\bigg)\bigg]-\frac{2T}{b}\psi^{(0)}\bigg(\frac{a}{b}+1\bigg)\bigg\rbrace,
\end{align}

\noindent where we have substituted $N=\Omega/(2\pi T)$. The evaluation of the $\Omega$ limit can be achieved by resorting to the series expansion
\begin{equation}
\psi^{(0)}(z+1)=\ln(z)+\frac{1}{2z}-\sum^\infty_{\ell=1}\frac{B_{2\ell}}{2\ell z^{2\ell}},
\end{equation}
where the coefficients $B_{2\ell}$ denote the Bernoulli numbers. Applying this formula, the expression for $\mathcal{I}_1$ evaluates to
\begin{equation}\label{Eq_I1_Identity}
\mathcal{I}_1=\nu_0\int\frac{d^2\mathbf{k}}{(2\pi)^2}\bigg\lbrace \frac{T}{a}+\frac{2T}{b}\bigg[\ln\bigg(\frac{a}{b}\bigg)-\psi^{(0)}\bigg(\frac{a}{b}+1\bigg)\bigg]\bigg\rbrace.
\end{equation}
Finally, by substituting the expressions for $a$ and $b$ and then rescaling the momentum integral according to
\begin{equation}
k=\bigg(\frac{2\pi\mathcal{B}T}{v_F}\bigg)^{1/3}y,
\end{equation}
one finds that Eq. \eqref{Eq_I1_Identity} simplifies to
\begin{equation}\label{Eq_I1_Final}
\mathcal{I}_1=\frac{\nu_0T}{2\pi}\Xi\bigg[\frac{m(T)}{(2\pi\mathcal{B}T/v_F)^{1/3}},\frac{\lambda^{1/2}_\text{latt}}{(2\pi\mathcal{B}T/v_F)^{1/3}}\bigg],
\end{equation}

\noindent where
\begin{widetext}
\begin{align}\label{Eq_Xi}
\Xi(x,\tau)=&\int^\infty_0 dy\int^{2\pi}_0\frac{d\theta}{2\pi}\bigg(\frac{y}{y^2+x^2+\tau^2\cos^2(2\theta)}+\frac{2y^2}{\cos^2(2\theta)}\bigg\lbrace\ln\bigg[\frac{y^3+[x^2+\tau^2\cos^2(2\theta)]y}{\cos^2(2\theta)}\bigg]\nonumber\\
&-\psi^{(0)}\bigg[\frac{y^3+[x^2+\tau^2\cos^2(2\theta)]y}{\cos^2(2\theta)}+1\bigg]\bigg\rbrace\bigg).
\end{align}

In order to determine $\mathcal{I}_2$, we expand the integrand on the right-hand side of Eq. \eqref{Eq_I2} in powers of $m(T)$. By performing the substitutions $k=\mathcal{B}\mathcal{C}^{-1/2}\rho$ and $\omega=v_F\mathcal{B}^2\mathcal{C}^{-3/2}v$, we arrive at the result
\begin{align}
\mathcal{I}_2=-\frac{\nu_0v_Fm^2(T)}{4\pi^3\sqrt{\mathcal{C}}}\int^{\Lambda\sqrt{\mathcal{C}}/\mathcal{B}}_0d\rho\int^{2\pi}_0d\theta\int^{\infty}_0dv\frac{\rho^5}{[\rho^4+4\kappa_\text{latt}\rho^2\cos^2(2\theta)+\rho v\cos^2(2\theta)+v^2\sin^2(2\theta)]^2},
\end{align}
\end{widetext}

\noindent where the dimensionless parameter $\kappa_\text{latt}=\lambda_\text{latt}/(\nu_0g^2_\text{nem})$ measures the strength of the nemato-elastic coupling with respect to the nematic interaction.
Upon substituting the expressions for $\mathcal{I}_1$ and $\mathcal{I}_2$ obtained here into Eq. \eqref{Eq_Nem_Mass_Def}, we arrive at the result given in Eq. \eqref{Eq_Nem_Mass}.

\section{Asymptotic solution of the nematic-mass equation}\label{Asymp_Mass_Sol}
\subsection{Low-temperature behavior}

This section is aimed at determining the asymptotic dependence of $m^2(T)$ on both $T$ and $\lambda_\text{latt}$ when the effective coupling
\begin{equation}
\Lambda_\text{latt}(T)=\frac{\lambda^{1/2}_\text{latt}}{(2\pi\mathcal{B}T/v_F)^{1/3}}
\end{equation}
becomes larger than any energy scale  of the model [see Eq. \eqref{Eq_Nem_Mass}]. In other words, this is equivalent to considering the temperature regime $T\ll\kappa^{3/2}_\text{latt}\varepsilon_F$, with $\varepsilon_F$ being the Fermi energy of the system.

\begin{figure}[t]
\centering \includegraphics[width=1.0\linewidth]{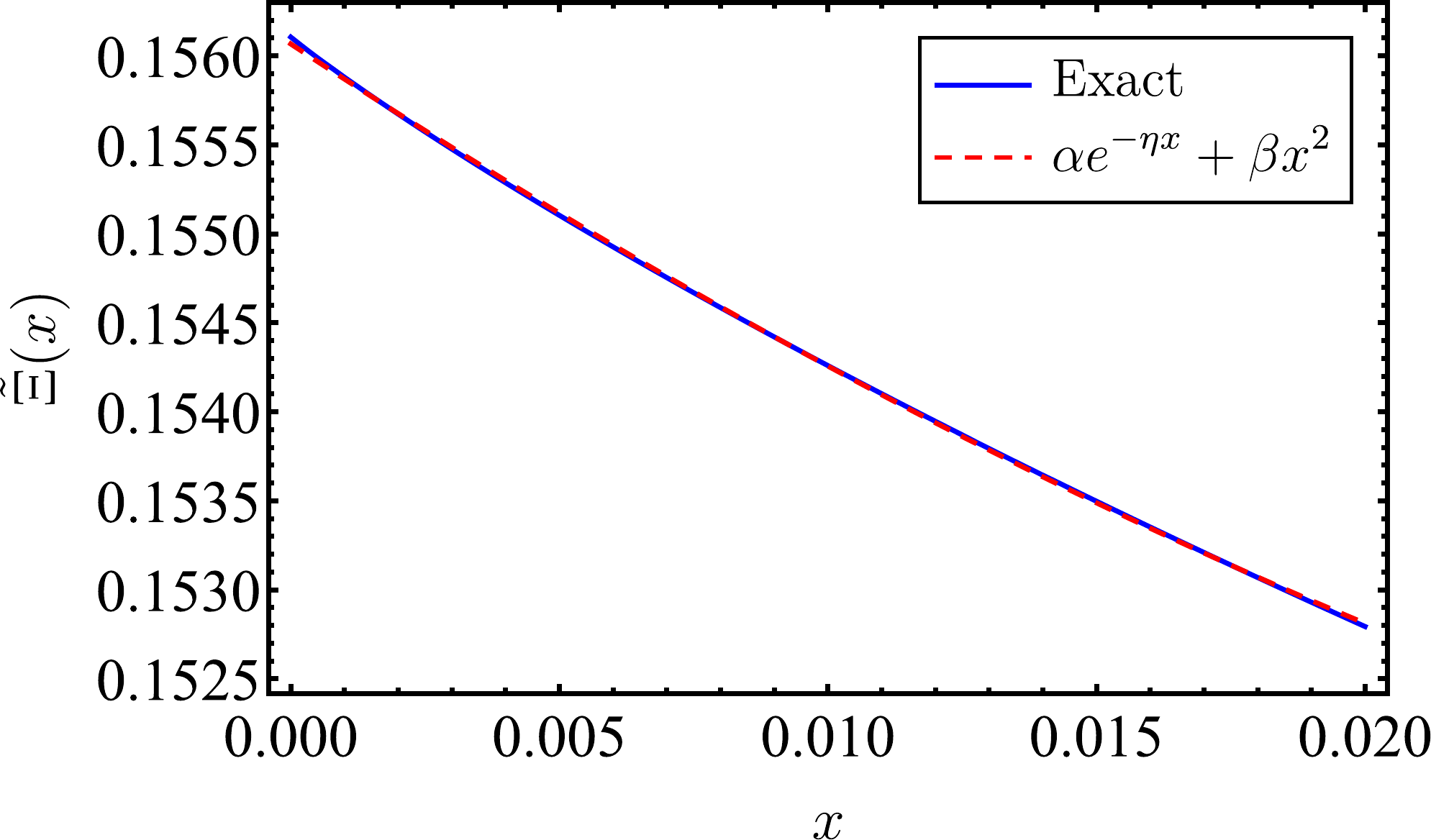}
\caption{Comparison between the dependence of both functions $\tilde{\Xi}(x)$ and the fit given by $f(x)=\alpha e^{-\eta x}+\beta x^2$ for the interval $0<x\ll 1$. The parameters in the latter function are given by $\alpha\approx0.156$, $\beta\approx1.698$, and $\eta\approx1.276$.}\label{Tilde_Xi_Dependence}
\end{figure}

To begin with, we linearize the cosine function $\cos(2\theta)$ on the right-hand side of Eq. \eqref{Eq_Xi} around the cold spots $\theta_n=(2n+1)\pi/4$. As a result, we are able to write
\begin{equation}
\int^{2\pi}_0\frac{d\theta}{2\pi}(\cdots)\approx\sum^{n_c}_{n=1}\int^{\theta_n+\theta_0}_{\theta_n-\theta_0}\frac{d\theta}{2\pi}(\cdots),
\end{equation}
where $n_c=4$ refers to the number of cold spots in the interval $[0,2\pi)$, and $\theta_0$ is an angular cutoff that sets the validity of the linear approximation for $\cos(2\theta)$. By performing the following substitutions
\begin{align}
y&\rightarrow\frac{y}{\Lambda^2_\text{latt}(T)},\\
\theta&\rightarrow\theta_n+\frac{z}{2\Lambda^3_\text{latt}(T)},
\end{align}
one obtains
\begin{widetext}
\begin{equation}\label{Eq_Rel_01}
\Xi\bigg[\frac{m(T)}{(2\pi\mathcal{B}T/v_F)^{1/3}},\frac{\lambda^{1/2}_\text{latt}}{(2\pi\mathcal{B}T)^{1/3}}\bigg]=n_c\bigg[\frac{\lambda^{1/2}_\text{latt}}{(2\pi\mathcal{B}T/v_F)^{1/3}}\bigg]^{-3}\tilde{\Xi}\bigg[\bigg(\frac{\lambda^{1/2}_\text{latt}}{(2\pi\mathcal{B}T/v_F)^{1/3}}\bigg)^2\frac{m(T)}{(2\pi\mathcal{B}T/v_F)^{1/3}}\bigg],
\end{equation}
where
\begin{align}
\tilde{\Xi}(x)=\frac{1}{2\pi}\int^\infty_0 dy\int^\infty_0dz\bigg(\frac{y}{x^2+y^2+z^2}+\frac{2y^2}{z^2}\bigg\lbrace\ln\bigg[\frac{y^3+(x^2+z^2)y}{z^2}\bigg]-\psi^{(0)}\bigg[\frac{y^3+(x^2+z^2)y}{z^2}+1\bigg]\bigg\rbrace\bigg).
\end{align}
\end{widetext}
Here, we have sent the upper cutoff $z_0=\frac{\theta_0v_F\lambda^{3/2}_\text{latt}}{\pi\mathcal{B}T}$ in the integration over the $z$ variable to infinity, because we are interested in the limit of lower temperatures. As a result, if we consider the interval $0<x\ll 1$, $\tilde{\Xi}(x)$ can be well approximated by
\begin{equation}\label{Eq_Rel_02}
\tilde{\Xi}(x)=\alpha e^{-\eta x}+\beta x^2,
\end{equation}
where the parameters on the right-hand side are given by $\alpha\approx0.156$, $\beta\approx1.698$, and $\eta\approx1.276$ (see Fig. \ref{Tilde_Xi_Dependence}).

Next, the substitution of Eqs. \eqref{Eq_Rel_01} and \eqref{Eq_Rel_02} into Eq. \eqref{Eq_Nem_Mass} yields
\begin{align}
&\bigg[1-\frac{\beta n_cv_F\nu_0U\lambda^{1/2}_\text{latt}}{4\pi^2(1+\gamma)\mathcal{B}N_f}\bigg]m^2(T)=\frac{\Delta s}{1+\gamma}+\frac{\alpha n_c\mathcal{B}\nu_0UT^2}{(1+\gamma)v_F\lambda^{3/2}_\text{latt}N_f}\nonumber\\
&\times\exp\bigg[-\frac{\eta v_F\lambda_\text{latt}}{2\pi\mathcal{B}T}m(T)\bigg].
\end{align}
At the nematic QCP, the above equation can be solved exactly. Indeed, by writing it as $m^2(T)=m_0\exp[-\kappa_0m(T)]$, its solution evaluates to 
\begin{equation}
m^2(T)=\frac{4}{\kappa^2_0}W^2_0\bigg(\frac{\sqrt{m_0}\kappa_0}{2}\bigg),
\end{equation}
with $W_0(z)$ being the principal branch of the so-called Lambert $W$ function, \textit{i.e.}, $W_0(0)=0$. By employing the definitions of $m_0$ and $\kappa_0$ in terms of the original model parameters, the expression for the nematic mass becomes
\begin{equation}\label{Eq_LowT_Nem_Mass}
m^2(T)=\frac{16\pi^2\mathcal{B}^2}{\eta^2v^2_F\lambda^2_\text{latt}}W^2_0\left(\frac{\eta}{2}\sqrt{\frac{\alpha\Gamma}{1-\beta\Gamma}}\right)T^2,
\end{equation}
where we have set
\begin{equation}\label{Eq_Gamma_Par}
\Gamma\equiv\frac{n_cv_F\nu_0U\lambda^{1/2}_\text{latt}}{4\pi^2(1+\gamma)\mathcal{B}N_f}.
\end{equation}
By making the substitutions $\mathcal{B}=\nu_0g^2_\text{nem}$ and $\varepsilon_F=v_Fk_F$ into Eq. \eqref{Eq_LowT_Nem_Mass}, one finds the expression for $m^2(T)$ given in the main text [see Eq. \eqref{Eq_MT_Nem_Mass}].

\subsection{High-temperature behavior}

Now we turn our attention to determining the asymptotic behavior of $m^2(T)$ for the interval of high temperatures $\kappa^{3/2}_\text{latt}\varepsilon_F\ll T\ll\varepsilon_F$. Since in this regime the nemato-elastic coupling becomes irrelevant, Eq. \eqref{Eq_Nem_Mass} can be approximated as
\begin{equation}\label{Eq_No_Lambda_Nem_Mass}
m^2(T)=\frac{\Delta s}{1+\gamma}+\frac{\nu_0UT}{2\pi(1+\gamma)N_f}\Xi\bigg[\frac{m(T)}{(2\pi\mathcal{B}T/v_F)^{1/3}},0\bigg],
\end{equation}
\noindent which is equivalent to the nematic-mass equation derived in Ref. \cite{Hartnoll-PRB(2014)}. To compare the lattice-free solution of Eq. \eqref{Eq_No_Lambda_Nem_Mass} with the expression for $m^2(T)$ obtained in Eq. \eqref{Eq_LowT_Nem_Mass}, we first notice that $\Xi(x,0)$ has the asymptotic form
\begin{equation}
\Xi(x,0)=\ln(1/x)-1.0747,
\end{equation}
as $x\rightarrow 0$. So within logarithmic accuracy and also considering that the system is exactly at its nematic QCP, we are able to write Eq. \eqref{Eq_No_Lambda_Nem_Mass} as
\begin{equation}
m^2(T)=\frac{\nu_0UT}{2\pi(1+\gamma)N_f}\ln\bigg[\frac{(2\pi\mathcal{B}T/v_F)^{1/3}}{m(T)}\bigg].
\end{equation}

The solution of the above equation can be expressed as
\begin{equation}
m^2(T)=\tilde{\kappa}^2_0\exp\bigg[-W_0\bigg(\frac{2\tilde{\kappa}^2_0}{\tilde{m}_0}\bigg)\bigg],
\end{equation}
where
\begin{align}
\tilde{m}_0&\equiv\frac{\nu_0UT}{2\pi(1+\gamma)N_f},\\
\tilde{\kappa}_0&\equiv\bigg(\frac{2\pi\mathcal{B}T}{v_F}\bigg)^{1/3}.
\end{align}
Since for $x\gg 1$, the Lambert function $W_0(x)$ can be well described to first order approximation by
\begin{equation}
W_0(x)\approx\ln\bigg(\frac{x}{\ln x}\bigg),
\end{equation}
we come to the conclusion that the mass of the nematic fluctuations for $\kappa^{3/2}_\text{latt}\varepsilon_F\ll T\ll\varepsilon_F$ is asymptotically given by
\begin{equation}
m^2(T)=\frac{k^2_FU}{12\pi^2(1+\gamma)N_f}\frac{T}{\varepsilon_F}\ln\bigg[\frac{256\pi^6(1+\gamma)^3g^4_\text{nem}N^3_f}{U^3\varepsilon^2_F}\frac{\varepsilon_F}{T}\bigg],
\end{equation}
where we have made the substitutions $\mathcal{B}=\nu_0g^2_\text{nem}$ and $\varepsilon_F=v_Fk_F$.

\section{Susceptibility $\chi_{JP}$}\label{Chi_JP}

\begin{figure}[t]
\centering
\includegraphics[width=0.80\linewidth]{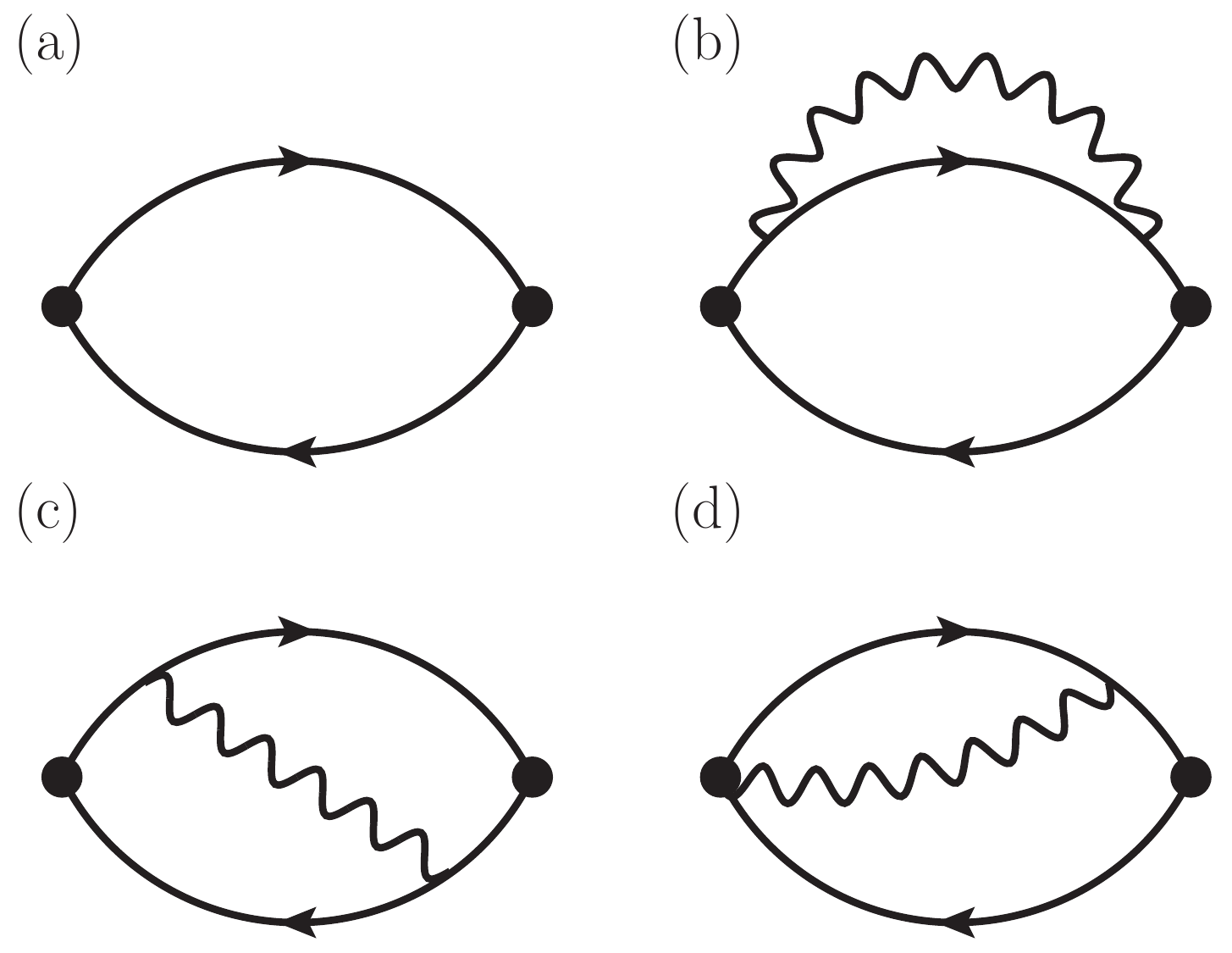}
\caption{Some Feynman diagrams for the calculation of the susceptibility $\chi_{JP}(T)$. For conciseness, we display here only the diagrams that yield a finite result in this calculation.}\label{Susceptibility_Diag}
\end{figure}

In this appendix, we calculate the temperature dependence of the susceptibility $\chi_{JP}(T)$. To do that, we will follow a similar treatment as discussed in Ref. \cite{Hartnoll-PRB(2014)} for a lattice-free model. For conciseness, only the Feynman diagrams that yield a finite result in this calculation are displayed in Fig. \ref{Susceptibility_Diag}. The remaining Feynman diagrams contain fermionic tadpoles and, as a consequence, they turn out to vanish in the present theory.

The first diagram represented in Fig. \ref{Susceptibility_Diag}(a) readily evaluates to
\begin{align}
\mathcal{D}_a&=-T\sum_{\nu_n}\int\frac{d^2 \mathbf{q}}{(2\pi)^2} \frac{q_x^2}{2m_f} G_{0}(\mathbf{k+q},i\omega_n+i\nu_n)\nonumber\\
&\times G_{0}(\mathbf{q},i\nu_n)\nonumber\\
&=-\int\frac{d^2 \mathbf{q}}{(2\pi)^2} \frac{q_x^2}{2m_f}\frac{[n_F(\xi_{\mathbf{q+k/2}})-n_F(\xi_{\mathbf{q-k/2}})]}{-i\omega_n+\xi_{\mathbf{q+k/2}}-\xi_{\mathbf{q-k/2}}}\nonumber\\
&\approx-\int\frac{d^2 \mathbf{q}}{(2\pi)^2} \frac{q_x^2}{2m_f} n'_F(\xi_{\mathbf{q}})\equiv\chi_0,
\end{align}
where $n_{F}(\omega)=1/(e^{\omega/T}+1)$ is the Fermi distribution function. The above integral naturally yields a constant contribution (denoted by $\chi_0$) to $\chi_{JP}(T)$. 

As for the diagram represented in Fig. \ref{Susceptibility_Diag}(b), we obtain
\begin{align}
\mathcal{D}_b&=2T\sum_{\Omega_m}\int\frac{d^2 \mathbf{k}}{(2\pi)^2}\Sigma_1(\mathbf{k},i\Omega_m)D(\mathbf{k},i\Omega_m),
\end{align}
where $\Sigma_1$ is a bosonic self-energy on one fermion line (the factor 2 is related to the fact that there are two diagrams that give exactly the same result). The corresponding bosonic self-energy $\Sigma_1$ is given by
\begin{align}
\Sigma_1=&-\frac{g_{\text{nem}}^2}{\nu_0 m_f}T\sum_{\nu_n}\int\frac{d^2 \mathbf{q}}{(2\pi)^2}q_x^2 V_{\mathbf{k+q,q}}^2 G_{0}(\mathbf{k+q},i\Omega_m+i\nu_n)\nonumber\\
&\times G_{0}^3(\mathbf{q},i\nu_n).
\end{align}
Since we are working with the continuum theory described by Eq. \eqref{Lagrangian}, the $d$-wave form factor may be written as $V_{\mathbf{p,q}}=-(p_x^2-p_y^2+q_x^2-q_y^2)$. In addition, for the diagram in Fig. \ref{Susceptibility_Diag}(c), we have
\begin{align}
\mathcal{D}_c&=T\sum_{\Omega_m}\int\frac{d^2 \mathbf{k}}{(2\pi)^2}\Sigma_2(\mathbf{k},i\Omega_m)D(\mathbf{k},i\Omega_m),
\end{align}
where we defined the quantity $\Sigma_2$ (despite the notation, this is not a self-energy insertion), which is given by
\begin{align}
\Sigma_2&=-\frac{g_{\text{nem}}^2}{\nu_0 m_f}T\sum_{\nu_n}\int\frac{d^2 \mathbf{q}}{(2\pi)^2}q_x (q_x+k_x) V_{\mathbf{k+q,q}}^2 \nonumber\\
&\times G_{0}^2(\mathbf{k+q},i\Omega_m+i\nu_n)G_{0}^2(\mathbf{q},i\nu_n)\nonumber\\
&=\Sigma_2^{(a)}+\Sigma_2^{(b)},
\end{align}
where $\Sigma_2^{(a)}$ is the term in the integral proportional to $q_x^2$ and $\Sigma_2^{(b)}$ is the term in the integral proportional to $q_x k_x$. In a similar way, for the diagram in Fig. \ref{Susceptibility_Diag}(d), we obtain
\begin{align}
\mathcal{D}_d&=T\sum_{\Omega_m}\int\frac{d^2 \mathbf{k}}{(2\pi)^2}\Sigma_3(\mathbf{k},i\Omega_m)D(\mathbf{k},i\Omega_m),
\end{align}
where we defined $\Sigma_3$, which is given by
\begin{align}
\Sigma_3&=-8\frac{g_{\text{nem}}^2}{\nu_0} T\sum_{\nu_n}\int\frac{d^2 \mathbf{q}}{(2\pi)^2}q_x \left(q_x+\frac{k_x}{2}\right) V_{\mathbf{k+q,q}} \nonumber\\
&\times G_{0}(\mathbf{k+q},i\Omega_m+i\nu_n)G_{0}^2(\mathbf{q},i\nu_n).
\end{align}

Collecting all the above terms, it can be shown that  $\Sigma_1+\Sigma_2+\Sigma_3\equiv\mathcal{C}_0$, where $\mathcal{C}_0$ is a constant given approximately by
\begin{align}
\mathcal{C}_0&\approx -\frac{g_{\text{nem}}^2}{\nu_0} \int\frac{d^2 \mathbf{q}}{(2\pi)^2}\bigg[\frac{q_x^2}{2m_f} V_{\mathbf{q,q}}^2 n'''_F(\xi_{\mathbf{q}})+8q_x^2 V_{\mathbf{q,q}}n''_F(\xi_{\mathbf{q}})\bigg].
\end{align}
\noindent Therefore, we can write
\begin{equation}\label{Eq_Temp_Chi_JP}
\chi_{JP}(T)= \chi_0-\mathcal{C}_0T\sum_{\Omega_m}\int\frac{d^2 \mathbf{k}}{(2\pi)^2}D(\mathbf{k},i\Omega_m),
\end{equation}
where $\chi_0$ and $\mathcal{C}_0$ are temperature-independent constants and $D(\mathbf{k},i\Omega_m)$ is the bosonic propagator given by Eq. \eqref{Eq_Prop_Temp}. By subtracting from Eq. \eqref{Eq_Temp_Chi_JP} the zero-temperature susceptibility $\chi^{(0)}_{JP}=\chi_{JP}(T\rightarrow 0)$ at $s=s_c$ and then employing the Eq. \eqref{Eq_Nem_Mass_Def} for the mass of the nematic fluctuations, we finally obtain 
\begin{equation}
\chi_{JP}(T)=\chi^{(0)}_{JP}-\frac{\mathcal{C}_0N_f}{U}[m^2(T)-\Delta s].
\end{equation}
As a result, if we consider the temperature regime $T\ll\kappa^{3/2}_\text{latt}\varepsilon_F$ and also suppose that the system is at its nematic QCP (see Appendix \ref{Asymp_Mass_Sol}), the susceptibility $\chi_{JP}(T)$ evaluates to
\begin{equation}
\chi_{JP}(T)=\chi^{(0)}_{JP}-\frac{16\pi^2\mathcal{C}_0N_fk^2_F}{\eta^2U\kappa^2_\text{latt}}W^2_0\left(\frac{\eta}{2}\sqrt{\frac{\alpha\Gamma}{1-\beta\Gamma}}\right)\bigg(\frac{T}{\varepsilon_F}\bigg)^2.
\end{equation}
On the other hand, for the interval of temperature $\kappa^{3/2}_\text{latt}\varepsilon_F\ll T\ll\varepsilon_F$ the nemato-elastic coupling can be dropped and, as a result, the susceptibility $\chi_{JP}(T)$ becomes
\begin{align}
\chi_{JP}(T)&=\chi^{(0)}_{JP}-\frac{\mathcal{C}_0k^2_F}{12\pi^2(1+\gamma)}\frac{T}{\varepsilon_F}\nonumber\\
&\times\ln\bigg[\frac{256\pi^6(1+\gamma)^3g^4_\text{nem}N^3_f}{U^3\varepsilon^2_F}\frac{\varepsilon_F}{T}\bigg].
\end{align}
The last two results are the limits of $\chi_{JP}(T)$ discussed in the main text.

\section{Approximate estimate of the resistivity due to random-field disorder using the fermionic Green's function}\label{Fermionic_Greens_function}

\begin{figure}[t]\label{Figure_Comparison}
\centering \includegraphics[width=0.95\linewidth]{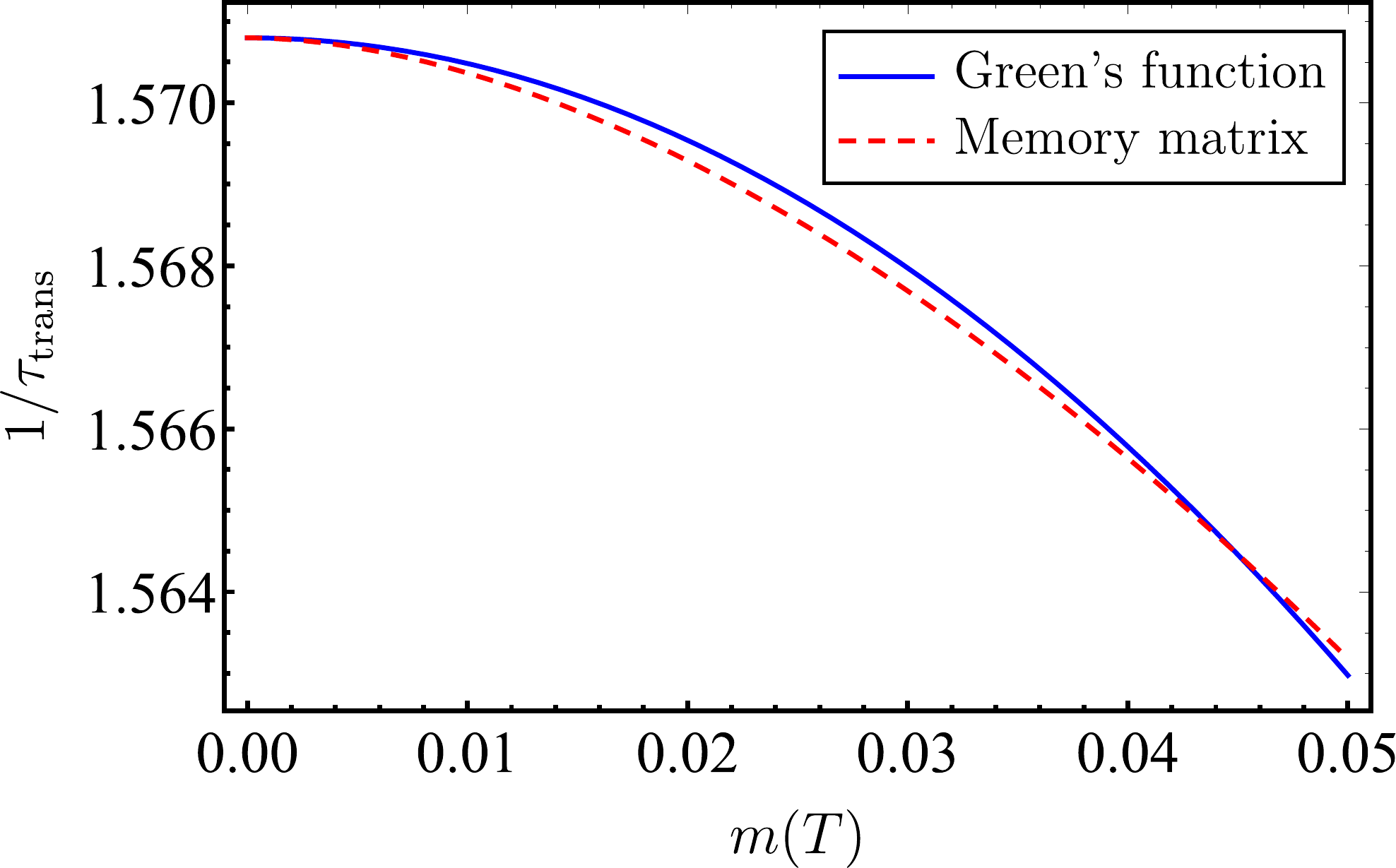}
\caption{Behavior of the inverse of the transport scattering time $\tau_\text{trans}$ as a function of $m(T)$ according to the memory matrix approach derived in our Eq. \eqref{Eq_Exac_Res} and a traditional Green's function method (for simplicity, we set $\lambda_{\text{latt}}=1$). We note that the use of the fermionic  Green's function approach here should not be viewed as a perturbatively exact result. This latter calculation only provides an approximate estimate and a further confirmation of our result in Eq. \eqref{Eq_Exac_Res}.}\label{Figure_Comparison}
\end{figure}

As explained in the main text, in order to further confirm our result obtained previously in Eq. \eqref{Eq_Exac_Res} using the memory matrix approach, we provide in this appendix an alternative (but approximate) way to calculate the resistivity due to random-field disorder using a fermionic Green's function method. In this respect, consider, e.g., a fermion in the vicinity of the Fermi surface constructed in terms of Fermi points given, without loss of generality, by $(k_F,0)$. Next, we will represent its deviation from these Fermi points by the momentum $\mathbf{q}$. The interacting fermionic Green's function will then be given by 
\begin{equation}
G(\mathbf{q},i\omega_n)=\frac{1}{i\omega_n-v_F q_x-q_y^2-\Sigma(\mathbf{q},i\omega_n)},
\end{equation}
where $\Sigma(\mathbf{q},i\omega_n)$ is the corresponding self-energy. At the Fermi momentum $k_F$, the lowest-order contribution to the self-energy after impurity-averaging procedure (which was explained in the main text) due to random-field disorder effects becomes 
\begin{align}\label{self_energy_impurity}
\Sigma_\text{\text{random}}(i\omega_n)&=\parbox{4.0cm}{\includegraphics[width=0.9 \linewidth]{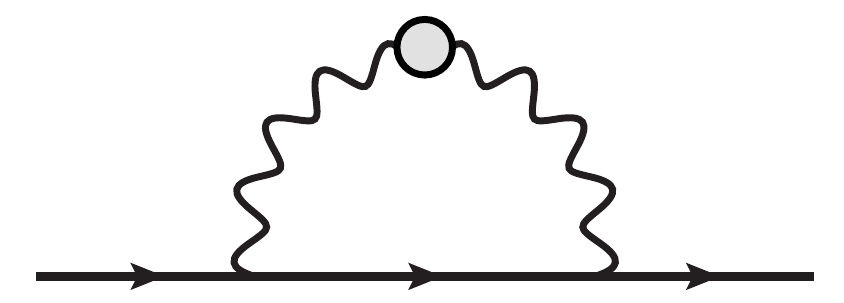}}\nonumber\\
&=h^2_0\int\frac{d^2\mathbf{q}}{(2\pi)^2}D^2(\mathbf{q},i\omega_n=0)\nonumber\\
&\times\frac{1}{i\omega_n-v_Fq_x-q^2_y-\Sigma(\mathbf{q},i\omega_n)},
\end{align}
where the circle in the above Feynman diagram represents the coupling $h^2_0$ of the order-parameter bosons with the random-field disorder. Since we can write $q_y=k_F\theta$, by using a trigonometric identity and the fact that $\sin\theta\approx \theta$, we obtain the following result $\cos^2(2\theta)\approx[1-(q_y/k_F)^2]^2$. 

To make further progress analytically, we will consider here an approximation similar to the one used in Refs. \cite{Metlitski-PRBa(2010),Hartnoll-PRB(2014)} and neglect the $q_x$-dependence on both the bosonic propagator $D$ and the fermionic self-energy $\Sigma$. As a consequence of this, we can now perform the integration over $q_x$ in Eq. \eqref{self_energy_impurity} and, by using the following identity
\begin{equation}
\text{sgn}(\omega)=-\frac{i}{\pi}\int_{-\infty}^{\infty}\frac{dx}{x-i\omega},
\end{equation}
we arrive at the result 
\begin{equation}
\Sigma_{\text{random}}(i\omega_n)=-i\text{sgn}(\omega_n)\frac{h_0^2}{4\pi v_F}\int_{-\infty}^{\infty} dq_y D^2(q_y,i\omega_n=0).
\end{equation}
The above expression represents the fermionic scattering rate given by $(1/\tau)$. However, in order to calculate the transport scattering time given by $\tau_{\text{trans}}$ that appears in the transport properties, we have to include an additional factor given by $(1-\cos\theta)$, which takes into account the fact that small-angle scattering does not relax the electric current in the model. This factor will add a contribution to the above integral that we can approximate as $(1-\cos\theta)\approx q_y^2/2 k_F^2$. Therefore, we obtain that the inverse of the scattering time $(1/\tau_{\text{trans}})$ is given by 
\begin{equation}
\frac{1}{\tau_{\text{trans}}}\propto h_0^2\int_{-\infty}^{\infty}dq_y\frac{q_y^2}{[q_y^2+{\lambda}_{\text{latt}}(1-q_y^2/k^2_F)^2+m^2(T)]^2}.
\end{equation}
Since the analytical solution of the above integral assumes a cumbersome form, we will focus only on its numerical result displayed in Fig. \ref{Figure_Comparison}. We observe from this plot that this approximate estimate of the inverse of the scattering time $(1/\tau_{\text{trans}})$ using a traditional Green's function method agrees well with our result in Eq. \eqref{Eq_Exac_Res} for small $m(T)$, thus confirming our result obtained within the memory matrix approach in the main text.


%

\end{document}